\documentclass{cmi}
\usepackage{graphicx}
\usepackage{url}
\usepackage{amssymb}
\usepackage{fancyvrb}
\usepackage{dsfont}
\usepackage{amsmath}
\usepackage{amssymb}
\usepackage{listings}
\usepackage{marvosym}
\usepackage{multirow}
\usepackage{hyperref}
\usepackage{algorithm}
\usepackage[noend]{algpseudocode}
\usepackage{algpseudocode}
\usepackage{colortbl}

\makeatletter
\@addtoreset{algorithm}{chapter}
\makeatother

\floatname{algorithm}{Алгоритм}

\DeclareGraphicsExtensions{.pdf}
\graphicspath{{pic/},{pic/pdf/}}

\begin{document}
	
\classify{УДК 004.272.25, 004.421, 004.032.24}



\newcolumntype{M}[1]{>{\centering\arraybackslash}m{#1}}
\newcolumntype{N}{@{}m{0pt}@{}}

\title{СОВМЕСТНОЕ ИСПОЛЬЗОВАНИЕ ТЕХНОЛОГИЙ MPI и OpenMP \\
ДЛЯ ПАРАЛЛЕЛЬНОГО ПОИСКА ПОХОЖИХ ПОДПОСЛЕДОВАТЕЛЬНОСТЕЙ \\
В СВЕРХБОЛЬШИХ ВРЕМЕННЫХ РЯДАХ НА ВЫЧИСЛИТЕЛЬНОМ КЛАСТЕРЕ \\
С УЗЛАМИ НА БАЗЕ МНОГОЯДЕРНЫХ ПРОЦЕССОРОВ \\
INTEL XEON PHI KNIGHTS LANDING}

\author{Я.А.~Краева\footnote{Южно-Уральский государственный университет (национальный исследовательский университет), кафедра системного программирования, просп. Ленина, 76, 454080, Челябинск; магистрант, e-mail: \href{mailto:kraevaya@susu.ru}{kraevaya@susu.ru}}, М.Л.~Цымблер\footnote{Южно-Уральский государственный университет (национальный исследовательский университет), кафедра системного программирования, просп. Ленина, 76, 454080, Челябинск; к.ф.-м.н., доцент, e-mail: \href{mailto:mzym@susu.ru}{mzym@susu.ru}}}


\maketitle{}

\begin{abstract}
В настоящее время поиск похожих подпоследовательностей требуется в широком спектре приложений интеллектуального анализа временных рядов: моделирование климата, финансовые прогнозы, медицинские исследования и др. В большинстве указанных приложений при поиске используется мера схожести Dynamic Time Warping (DTW), поскольку на сегодня научное сообщество признает меру DTW одной из лучших для большинства предметных областей. Мера DTW имеет квадратичную вычислительную сложность относительно длины искомой подпоследовательности, в силу чего разработан ряд параллельных алгоритмов ее вычисления на устройствах FPGA и многоядерных ускорителях с архитектурами GPU и Intel MIC. В данной статье предлагается новый параллельный алгоритм для поиска похожих подпоследовательностей в сверхбольших временных рядах на кластерных системах с узлами на базе многоядерных процессоров Intel Xeon Phi поколения Knights Landing (KNL). Вычисления распараллеливаются на двух уровнях: на уровне всех узлов кластера~--- с помощью технологии MPI, в рамках одного узла кластера~--- с помощью технологии OpenMP. Алгоритм предполагает использование дополнительных структур данных и избыточных вычислений, позволяющих эффективно задействовать возможности векторизации вычислений на процессорных системах Phi KNL. Эксперименты, проведенные на синтетических и реальных наборах данных, показали хорошую масштабируемость алгоритма.

\end{abstract}


\markboth{Я.А.~Краева, М.Л.~Цымблер}{Совместное использование технологий MPI и OpenMP для параллельного поиска\ldots}

\section{Введение.}
\label{sec:Introduction}

В настоящее время поиск похожих подпоследовательностей требуется в широком спектре приложений интеллектуального анализа временных рядов: моделирование климата, финансовые прогнозы, медицинские исследования и др. Задача поиска похожей подпоследовательности неформально может быть определена следующим образом. Пусть имеется временной ряд и поисковый запрос (временной ряд, длина которого существенно меньшей длины исходного ряда). Необходимо найти подпоследовательность исходного ряда, которая максимально похожа по форме на поисковый запрос.

На сегодня мера \emph{DTW (Dynamic Time Warping, динамическая трансформация временной шкалы)}~\cite{DBLP:conf/kdd/BerndtC94} признается научным сообществом лучшей мерой схожести временных рядов для многих предметных областей~\cite{DBLP:journals/pvldb/DingTSWK08}, поскольку позволяет адекватно измерять схожесть временных рядов, которые имеют различные длины либо сдвиг, растяжение или сжатие друг относительно друга. Поскольку мера \emph{DTW} имеет квадратичную вычислительную сложность относительно длины поискового запроса, были предложены методы снижения вычислительной нагрузки: индексирование~\cite{keogh}, заблаговременный отказ от избыточных вычислений, повторное использование результатов ранее выполненных вычислений~\cite{sakurai} и др. Последовательный алгоритм \emph{UCR-DTW}~\cite{rakthanmanon}, интегрирующий множество указанных техник, является на сегодня, вероятно, самым быстрым последовательным алгоритмом поиска похожих подпоследовательностей~\cite{sart}. В то же время предложены параллельные алгоритмы поиска похожих подпоследовательностей для устройств FPGA~\cite{wang,zhang}, многоядерных ускорителей архитектур GPU~\cite{sart,zhang} и Intel Many Integrated Core (MIC)~\cite{zymbler}.

Рассматриваемые в данной статье многоядерные системы архитектуры Intel MIC являются конкурентоспособной альтернативой более распространенным системам NVIDIA GPU и FPGA. Системы Intel MIC основаны на распространенной архитектуре Intel x86 и поддерживают соответствующие модели и инструменты параллельного программирования. Устройства MIC обеспечивают большое количество энергоэффективных вычислительных ядер с малой (относительно обычных многоядерных процессоров Intel) тактовой частотой, обладающих высокой пропускной способностью локальной памяти и 512-битными векторными регистрами. В настоящее время компания Intel предлагает два поколения MIC-устройств, имеющих базовое имя Xeon Phi: сопроцессор Knights Corner (KNC)~\cite{chrysos} с 57--61~ядрами и самостоятельная процессорная система Knights Landing (KNL)~\cite{sodani} с 64--72~ядрами соответственно. Системы MIC, как правило, дают наибольшую производительность в таких приложениях, где имеет место большой объем данных, вовлеченных в вычисления (десятки миллионов элементов и более), которые могут быть векторизованы компилятором~\cite{SokolinskayaS16}. Под векторизацией понимается способность компилятора заменить несколько скалярных операций в теле цикла с фиксированным количеством повторений в одну векторную операцию~\cite{bacon}. 

Параллельный поиск похожих подпоследовательностей в сверхбольших временных рядах (порядка сотен миллиардов точек) на кластерных системах рассмотрен в работах~\cite{DBLP:conf/icacci/ShabibNNDPSATS15,MovchanZ16}. В статье~\cite{DBLP:conf/icacci/ShabibNNDPSATS15} предложен способ 
распараллеливания последовательного алгоритма поиска похожих подпоследовательностей \emph{UCR-DTW} для вычислительного кластера на основе использования фреймворка Apache Spark. В рамках данного фреймворка приложение запускается как задача, координируемая мастер-узлом вычислительного кластера, на котором установлен соответствующий драйвер. Алгоритм предполагает фрагментацию временного ряда, однако количество фрагментов не совпадает
с количеством вычислительных узлов кластерной системы. Фрагменты сохраняются в виде отдельных файлов, доступных всем узлам в силу использования распределенной файловой системы HDFS (Hadoop Distributed File System). Каждый фрагмент обрабатывается отдельной нитью, реализующей последовательный алгоритм \emph{UCR-DTW}. Количество процессов, запускаемых на узле кластера, равно количеству ядер процессора на данном узле. В экспериментах разработанный алгоритм опережает \emph{UCR-DTW} в 2--5~раз, однако представленные результаты его запусков на кластере из 2--6~узлов не позволяют говорить о хорошей масштабируемости решения.

В предыдущих исследованиях авторов данной статьи~\cite{MovchanZ16,zymbler} разработан параллельный алгоритм поиска похожих подпоследовательностей на вычислительном кластере с узлами на базе сопроцессоров Phi KNC. Предложенная вычислительная схема <<CPU+Phi>> предполагает фрагментацию ряда по узлам кластера и следующее разделение функций между процессором и KNC внутри узла. KNC выполняет вычисление меры \emph{DTW} для подпоследовательностей, загруженных в его память процессором, и выбирает подпоследовательность, максимально похожую на образец поиска. Процессор исключает из обработки заведомо непохожие на поисковый запрос подпоследовательности на основе вычисления нижних оценок формирует пакеты подпоследовательностей для выгрузки данных на сопроцессор. Предложенная схема вычислений обеспечивает высокую производительность, однако слабо задействует возможности векторизации вычислений Phi KNC. Масштабируемость, близкая к линейной, достигается алгоритмом в случае, когда поисковый запрос имеет длину от 4000~точек, что относительно редко встречается на практике. Помимо этого, данная схема не может быть применена для случая кластерной системы на базе узлов с системами следующего поколения Phi, поскольку KNL является независимым многоядерным устройством, запускающим приложения только в нативном режиме. В работе~\cite{KraevaZ18} авторами предложен подход к эффективному поиску похожих подпоследовательностей в оперативной памяти процессора Phi KNL. 

В данной статье предлагается новый параллельный алгоритм \emph{PhiBestMatch} для поиска похожих подпоследовательностей в сверхбольших временных рядах на кластерных системах с узлами на базе многоядерных процессоров Intel Xeon Phi поколения Knights Landing. Вычисления распараллеливаются на двух уровнях: на уровне всех узлов кластера~--- с помощью технологии MPI, в рамках одного узла кластера~--- с помощью технологии OpenMP. Алгоритм предполагает использование дополнительных структур данных и избыточных вычислений, позволяющих эффективно задействовать возможности векторизации вычислений на процессорных системах Phi KNL.

Статья организована следующим образом. В разделе~\ref{sec:Definition} приводится формальная постановка задачи и дано краткое описание последовательного алгоритма \emph{UCR-DTW}, используемого в качестве базиса нового параллельного алгоритма. Раздел~\ref{sec:Approach} содержит описание алгоритма \emph{PhiBestMatch}. В разделе~\ref{sec:Experiments} описаны вычислительные эксперименты, исследующие эффективность предложенного алгоритма. Заключение резюмирует результаты, полученные в рамках исследования.

\section{Постановка задачи.}
\label{sec:Definition}

\subsection{Формальные определения и нотация.}
\label{subsec:fotmalDefinition}

Дадим определения используемых терминов в соответствии с работой~\cite{rakthanmanon}. \textit{Временной ряд (time series)} $T$~--- это  хронологически упорядоченная последовательность вещественных значений: $T=(t_1, t_2,\dots,t_m)$, $t_i \in \mathbb{R}$. Число $m$ обозначается $\lvert T \rvert$ и называется длиной временного ряда.

\textit{Мерой схожести (similarity measure)} $\mathcal{D}$ между двумя временными рядами $X$ и $Y$ называется вещественная неотрицательная функция, вычисляющая расстояние между данными рядами: $\mathcal{D}(X,Y) \geqslant 0$.

Пусть даны два временных ряда $X = (x_{1}, x_{2},\dots,x_{m})$ и $Y = (y_{1}, y_{2},\dots,y_{m})$. Тогда \textit{динамической трансформацией временной шкалы (Dynamic Time Warping, DTW)} называется мера схожести $DTW(X,Y)$, вычисляемая следующим образом~\cite{DBLP:conf/kdd/BerndtC94}:

\begin{equation}
\label{dtw}
\begin{gathered}
DTW(X,Y) = d(m,m), \\
d(i,j) = (x_{i} - y_{j})^2 + \min
\begin{cases}d(i-1,j)\\
d(i, j-1)\\
d(i-1,j-1)\\
\end{cases}, \\
d(0,0) = 0,\;d(i,0) = d(0,j) = \infty,\;1\leqslant i \leqslant m,\;1\leqslant j \leqslant m. 
\end{gathered}
\end{equation}

Мера \emph{DTW} позволяет сравнивать ряды, которые смещены вдоль временной оси, сжаты, растянуты или имеют разные длины. Далее нами будет рассматриваться случай, когда временные ряды имеют одинаковые длины, поскольку это упрощает изложение и не ограничивает общность~\cite{rakthanmanon}.

В формуле~\eqref{dtw} матрица $(d_{ij}) \in \mathbb{R}^{m \times m}$ выражает соответствие между точками сравниваемых временных рядов и называется \emph{матрицей трансформации}.

\textit{Путь трансформации (warping path) P} представляет собой последовательность элементов матрицы трансформации, которая определяет соответствие между временными рядами $Q$ и $C$. Пусть элемент $p_t$ пути трансформации $P$ определяется как $p_t=(i,j)_t$, тогда 

\begin{equation}
\label{eq:WarpingPath}
P = p_{1},p_{2},\dots,p_{t},\dots,p_{T},\; m\leqslant T \leqslant 2n-1.	
\end{equation}

На путь трансформации накладывается ряд ограничений. Путь должен начинаться и заканчиваться в диагонально противоположных элементах матрицы трансформации, шаги пути ограничены соседними ячейками, а точки пути должны быть монотонно разнесены во времени.

Объем вычислений при подсчете меры схожести \emph{DTW} может быть уменьшен за счет огрубления схожести. Для этого на путь трансформации могут налагаться дополнительные ограничения. Одним из наиболее часто применяемых ограничений является т.н. \emph{полоса Сако---Чиба}~\cite{sakoe}, не позволяющая пути трансформации отклоняться более чем на $r$ элементов от диагонали матрицы трансформации. 

\textit{Подпоследовательностью (subsequence)} $T_{i,n}$ временного ряда $T$ называется непрерывное подмножество $T$, состоящее из $n$ элементов и начинающееся с позиции $i$: $T_{i,n} = (t_i, t_{i+1},\dots,t_{i+n-1}),\; 1 \leqslant i \leqslant m-n+1,\; k \ll m$. 

Определим решаемую в данной работе \textit{задачу поиска самой похожей подпоследовательности} ряда в смысле меры \emph{DTW}. Пусть имеется длинный временной ряд $T$ и заданный пользователем существенно более короткий временной ряд $Q$ (называемый \textit{поисковым запросом (query)}), где $m=\lvert T \rvert \gg \lvert Q \rvert = n$. Тогда \textit{самой похожей подпоследовательностью (best match)} назовем такую подпоследовательность $T_{i,n}$, форма которой максимально похожа на запрос $Q$ в смысле меры \emph{DTW}:

\begin{equation}
\exists i\; \forall k \; DTW(Q, T_{i,n}) \leqslant DTW(Q, T_{k,n}),\;1 \leqslant i,k \leqslant m-n+1.
\end{equation}

Далее для краткости изложения алгоритма подпоследовательность $T_{i,n}$ будем называть \textit{кандидатом} (на максимальную схожесть с запросом $Q$) и обозначать как $C$.

\subsection{Последовательный алгоритм.}
\label{subsec:UCR-DTW}

В настоящее время алгоритм \emph{UCR-DTW}~\cite{rakthanmanon} является одним из самых быстрых алгоритмов поиска похожих подпоследовательностей, интегрируя множество техник, ускоряющих поиск. Поскольку предлагаемый параллельный алгоритм использует \emph{UCR-DTW} в качестве основы, ниже кратко описаны его основные идеи.

\emph{Использование квадрата Евклидова расстояния.} \textit{Евклидово расстояние (ED)} между двумя временными рядами $Q$ и $C$, где $\lvert Q \rvert = \lvert C \rvert$, определяется следующим образом:
\begin{equation}
ED(Q,C)=\sqrt{\sum\limits_{i=1}^n(q_{i}-c_{i})^2}.
\label{Euclidean}
\end{equation}

Вычисление квадратного корня в \emph{ED}~\eqref{Euclidean} и \emph{DTW}~\eqref{dtw} можно опустить, т.к. это не изменит относительного ранжирования подпоследовательностей, похожих на запрос, поскольку обе функции монотонные и вогнутые.

\emph{Z-нормализация.} Перед вычислением меры схожести запрос и подпоследовательность временного ряда необходимо подвергнуть z-нормализации~\cite{tarango}. \textit{Z-нормализацией} временного ряда $T$ называется временной ряд $\hat{T}=(\hat{t}_{1},\dots,\hat{t}_{m})$, элементы которого вычисляются следующим образом:

\begin{equation}
\label{eq:Normalization1}
\begin{gathered}
\hat{t_i} = \frac{t_{i}-\mu}{\sigma},\; 1\leqslant i \leqslant m;\;\;	
\mu=\frac{1}{m}\sum\limits_{i=1}^mt_{i}; \;\;
\sigma^2=\frac{1}{m}\sum\limits_{i=1}^mt_i^2-\mu^2.
\end{gathered}
\end{equation}	

Z-нормализация позволяет сравнивать формы рядов, которые отличны по амплитуде. После нормализации среднее арифметическое временного ряда приблизительно равно 0, а среднеквадратичное отклонение близко к 1. 

\emph{Каскадное применение нижних границ схожести}. \emph{Нижняя граница схожести (lower bound, LB)} представляет собой функцию, вычислительная сложность которой меньше вычислительной сложности меры \emph{DTW}. Нижняя граница используется для отбрасывания кандидатов, заведомо непохожих на запрос, без вычисления меры \emph{DTW}~\cite{DBLP:journals/pvldb/DingTSWK08}. 

Обозначим за \textit{bsf (best-so-far)} лучшую текущую нижнюю оценку схожести текущей подпоследовательности $T_{i,n}$ и поискового запроса $Q$. Если нижняя граница для кандидата превышает порог \textit{bsf}, то значение меры \emph{DTW} для данного кандидата также превысит \textit{bsf}, и кандидат заведомо непохож на запрос. В процессе сканирования временного ряда алгоритм \emph{UCR-DTW} пытается улучшить (уменьшить) значение $bsf$. Значение \textit{bsf} инициализируется как $+\infty$ и на \textit{i}-м шаге поиска вычисляется следующим образом: 

\begin{equation}
\label{eq:bsf}
bsf_{(i)}= \operatorname{min}(bsf_{(i-1)}, \left \{
\begin{array}{l l}
+\infty &, LB(Q, T_{i,n}) > bsf_{(i-1)} \\
DTW(Q, T_{i,n}) &, otherwise
\end{array}
)\right.
\end{equation}

Алгоритм \emph{UCR-DTW} использует следующие нижние границы схожести: $LB_{Kim}FL$~\cite{rakthanmanon}, $LB_{Keogh}EC$ и $LB_{Keogh}EQ$~\cite{keogh}, которые применяются каскадным образом. 

Нижняя граница $LB_{Kim}FL$ представляет собой Евклидово расстояние между первой и последней парами точек $Q$ и $C$: 
\begin{equation}
\label{LB_Kim}
LB_{Kim}FL(Q,C) = ED(\hat{q}_1,\hat{c}_1) + ED(\hat{q}_n, \hat{c}_n).
\end{equation}	

Нижняя граница $LB_{Keogh}EC$ показывает минимальную схожесть между оболочкой запроса $E$ \textit{(envelope)} и кандидатом $\hat{C}$ и вычисляется следующим образом:

\begin{equation}
\label{eq:LBKeoghEC}
LB_{Keogh}EC(Q, C)=\sum_{i=1}^n
\begin{cases}
(\hat{c}_{i}-u_{i})^2,\; & if\; \hat{c}_{i}>u_{i}\\
(\hat{c}_{i}-\ell_{i})^2,\; & if\; \hat{c}_{i}<\ell_{i}\\
0, & otherwise\end{cases}.
\end{equation}

В формуле~\eqref{eq:LBKeoghEC} последовательности $U = (u_{1},..,u_{n})$ и $L = (\ell_{1},..,\ell_{n})$ обозначают \textit{верхнюю (upper)} и \textit{нижнюю (lower) границы оболочки} запроса $Q$, которые вычисляются по формуле~\eqref{envelope}:

\begin{equation}
\label{envelope}
\begin{gathered} 
u_{i}=\max(\hat{q}_{i-r},\dots,\hat{q}_{i+r}),\\
\ell_{i}=\min(\hat{q}_{i-r},\dots,\hat{q}_{i+r}).
\end{gathered}
\end{equation}

Нижняя граница $LB_{Keogh}EQ$ представляет собой Евклидово расстояние между запросом $Q$ и оболочкой кандидата $C$, то есть по сравнению с $LB_{Keogh}EC$ роли запроса и кандидата меняются местами:

\begin{equation}
\label{LBKeoghEQ}
LB_{Keogh}EQ(Q,C) := LB_{Keogh}EC(C,Q).	
\end{equation}


Выполнение алгоритма \emph{UCR-DTW} происходит следующим образом. Сначала нормализуется запрос и вычисляется его оболочка, порог $bsf$ инициализируется значением $\infty$. Далее алгоритм сканирует временной ряд от начала до конца, применяя к текущей подпоследовательности каскад нижних границ схожести. Если подпоследовательность не была отброшена, то вычисляется расстояние \emph{DTW}. Далее значение $bsf$ заменяется на вычисленное значение меры \emph{DTW}, если последнее меньше лучшей текущей оценки схожести. Таким образом, по окончании сканирования ряда алгоритм находит самую похожую подпоследовательность.

\section{Параллельный алгоритм PhiBestMatch.}
\label{sec:Approach}

В данном разделе представлен новый параллельный алгоритм поиска самой похожей подпоследовательности временного ряда \emph{PhiBestMatch} для кластерных систем с узлами на базе многоядерных процессорных систем архитектуры Intel MIC. Распараллеливание алгоритма основано на следующих основных принципах: параллелизм по данным, выравнивание данных в памяти, векторизация вычислений.

Параллелизм по данным \emph{на уровне вычислительных узлов кластерной системы} реализуется с помощью фрагментации временного ряда. Временной ряд разбивается на фрагменты (подпоследовательности) примерно равной длины, каждый из которых размещается в оперативной памяти отдельного  вычислительного узла, выполняющего обработку данного фрагмента. Данная схема предполагает следующее \emph{масштабирование}: если суммарный объем оперативной памяти недостаточен для размещения временного ряда, в кластер добавляется дополнительный узел (узлы).

Для обработки фрагмента на каждом узле запускается вычислительный процесс, и все процессы используют один и тот же алгоритм. В процессе обработки процессы обмениваются данными для сокращения объема вычислений. Обмены данными между процессами реализуются с помощью технологии MPI~\cite{DBLP:conf/pvm/Gropp12}. Параллелизм по данным \emph{внутри узла кластерной системы }реализуется следующим образом. В рамках вычислительного процесса на ядрах процессора узла запускаются (\emph{fork}) нити, разделяющие оперативную память узла. Нити осуществляют параллельную обработку фрагмента на базе технологии OpenMP~\cite{DBLP:conf/sc/Mattson06}. 

\emph{Векторизация циклов} является одним из ключевых условий достижения высокой производительности вычислительных программ на параллельных архитектурах~\cite{bacon}. Векторизация циклов заключается в преобразовании компилятором последовательности скалярных операторов из тела цикла в один векторный оператор. Таким образом, чтобы повысить производительность поиска похожих подпоследовательностей на Phi KNL, необходимо организовывать вычисления таким образом, чтобы увеличить в алгоритме, насколько это возможно, количество векторизуемых циклов.

Эффективность векторизации циклов, однако, может существенно снижена в силу не выровненного доступа к данным в оперативной памяти, который порождает эффект разделения цикла (loop peeling)~\cite{bacon}. Если начальный адрес массива не выровнен на ширину векторного регистра (т.е. количество элементов, которые могут быть загружены в векторный регистр), то компилятор разбивает цикл на три части. Первая часть итераций, которые обращаются к памяти с начального адреса до первого выровненного адреса, и третья часть итераций с последнего выровненного адреса до конечного адреса векторизуется отдельно.

В соответствии с этим в алгоритме \emph{PhiBestMatch} предлагается компоновка данных в оперативной памяти, которая обеспечивает выровненный доступ к подпоследовательностям временного ряда и соответствующая этой компоновке вычислительная схема, в которой вычисления реализованы в виде векторизуемых циклов.

\subsection{Фрагментация временного ряда.}
\label{subsec:fragmentation}

Фрагментация временного ряда обеспечивает параллелизм по данным на уровне вычислительных узлов кластерной системы и осуществляется следующим образом. Для предотвращения потери результирующих подпоследовательностей, находящихся на стыке фрагментов, предлагается техника \emph{разбиения с перекрытием}, которая заключается в следующем. В конец каждого фрагмента временного ряда, за исключением последнего по порядку, добавляется $n-1$ элементов ряда, взятых с начала следующего фрагмента, где $n$~--- длина запроса. Формальное определение разбиения с перекрытием выглядит следующим образом.

Пусть $N = \lvert T \rvert-n +1 = m-n+1$~--- количество подпоследовательностей, которые необходимо обработать, $F$~--- количество фрагментов, $T^{(k)}$~--- $k$-й фрагмент временного ряда $T=(t_1, t_2, \dots, t_m)$, где $0 \leqslant k \leqslant F-1$. Тогда \emph{фрагмент} $T^{(k)}$ определяется как подпоследовательность $T_{start, \, len}$, где 

\begin{equation} 
\label{eq:OverlapPartitioning}
\begin{gathered}
start = k \cdot \lfloor \tfrac{N}{F} \rfloor + 1 \\
len = \begin{cases}
\lfloor \tfrac{N}{F} \rfloor + (N\;mod\;F) + n - 1, & k=F-1 \\
\lfloor \tfrac{N}{F} \rfloor + n - 1, & otherwise.
\end{cases}
\end{gathered}
\end{equation} 

Количество фрагментов (вычислительных узлов кластерной системы) $F$ выбирается таким образом, чтобы фрагмент и соответствующие вспомогательные данные алгоритма могли быть размещены в оперативной памяти вычислительного узла.

\subsection{Компоновка данных.}
\label{subsec:dataLayout}

Предлагаемая компоновка данных в оперативной памяти вычислительного узла кластерной системы и обеспечивает представление фрагмента временного ряда и вспомогательных данных алгоритма в виде выровненных в памяти матриц, циклы обработки которых векторизуются компилятором.

Выравнивание данных выполняется следующим образом.
Пусть обработка подпоследовательности $T_{i,n}$ ряда $T$ осуществляется с использованием векторного регистра, вмещающего $w$ вещественных чисел. Если длина подпоследовательности не кратна $w$, то подпоследовательность дополняется фиктивными нулевыми элементами. Обозначим количество фиктивных элементов за $pad = w - (n\;mod\;w)$, тогда выровненная подпоследовательность $\tilde{T}_{i,n}$ определяется следующим образом:

\begin{equation}
\label{eq:alignedC}
\tilde{{T}}_{i,n} = \begin{cases}
t_{i}, t_{i+1},\dots,t_{i+n-1}, \underbrace{0, 0,\dots,0}_{pad}, & if\; n\;mod\;w > 0\\ 
t_{i}, t_{i+1},\dots,t_{i+n-1}, & otherwise.
\end{cases}
\end{equation}

Согласно определению~\eqref{dtw}, $\forall Q,\,C: DTW(Q,C) = DTW(\tilde{Q},\tilde{C})$. В дальнейшем изложении мы предполагаем, что запрос и подпоследовательности-кандидаты выровнены в памяти, и для упрощения используем обозначения и $Q$ и $C$ соответственно. Выравнивание подпоследовательностей позволяет избежать накладных расходов на разбивание циклов на несколько итераций (loop peeling).

Все (выровненные) подпоследовательности временного ряда сохраняются в виде матрицы подпоследовательностей для обеспечения векторизации вычислений.
\emph{Матрица подпоследовательностей} $S_{T}^n \in \mathbb {R}^{N\times(n+pad)}$ определяется следующим образом:

\begin{equation}
\label{eq:subsequenceMatrix}
S_{T}^n(i,j):=\tilde{t}_{i+j-1}.
\end{equation} 

Обозначим количество нижних границ схожести, используемых в алгоритме поиска самой похожей подпоследовательности, за $lb_{max}$ ($lb_{max} \geqslant 1$), и за $LB_1$, $LB_2$, $\dots$, $LB_{lb_{max}}$ обозначим эти границы, перечисленные в порядке их вычисления в каскаде применения нижних оценок. Тогда $L_T^n \in \mathbb{R}^{N \times lb_{max}}$, \emph{матрица нижних границ схожести} всех подпоследовательностей длины $n$ временного ряда $T$ с поисковым запросом $Q$, имеет следующий вид:

\begin{equation}
\label{eq:lbMatrix}
L_{T}^n(i,j):=LB_{j}(T_{i,n}, Q).
\end{equation}

\emph{Карта схожести} представляет собой вектор-столбец $B_T^n \in \mathbb {B}^{N}$, который для каждой подпоследовательности длины $n$ ряда $T$ хранит конъюнкцию применения всех нижних границ схожести этой подпоследовательности с текущим значением порога (\emph{bsf}):

\begin{equation}
\label{eq:bitmapMatrix}
B_{T}^n(i):=\bigwedge_{j=1}^{lb_{max}} (L_{T}^n(i,j) < bsf).
\end{equation} 

Введем матрицу для хранения подпоследовательностей-кандидатов, то есть тех подпоследовательностей из матрицы $S_T^n$, которые не были отброшены при применении нижней границы схожести. Данная матрица будет обрабатываться параллельно для вычисления значений меры \emph{DTW} между кандидатами и запросом. Далее минимальное из вычисленных значение меры схожести \emph{DTW} используется в качестве порога \emph{bsf}.

Параллельная обработка матрицы кандидатов использует разбиение строк этой матрицы на сегменты, обрабатываемые отдельными нитями одного узла кластера.
Обозначим число нитей, используемых параллельным алгоритмом, за $p\;(p\geqslant1)$. Введем \emph{размер сегмента} $s\in \mathbb{N}$ ($s \leqslant \lceil\frac{N}{p}\rceil$)~--- количество строк матрицы кандидатов, обрабатываемых одной нитью. Тогда \emph{матрица кандидатов} $C_T ^ n \in \mathbb{R} ^ {(s\cdot p) \times (n + pad)}$ определяется следующим образом:

\begin{equation}
\label{eq:candidateMatrix}
C_{T}^n(i,\cdot):=S_{T}^n(k,\cdot): B_{T}^n(i)=\texttt{TRUE}.
\end{equation} 

Несмотря на то, что вычисление матрицы кандидатов является распараллеливаемой операцией, векторизация операторов соответствующего цикла затруднена, поскольку в соответствии с определением~\eqref{dtw} при вычислении меры схожести \emph{DTW} имеет место зависимость по данным.

\subsection{Вычислительная схема алгоритма.}
\label{subsec:computationalScheme}

\begin{figure}[!ht]
	\center		
	\includegraphics[width=0.9\linewidth]{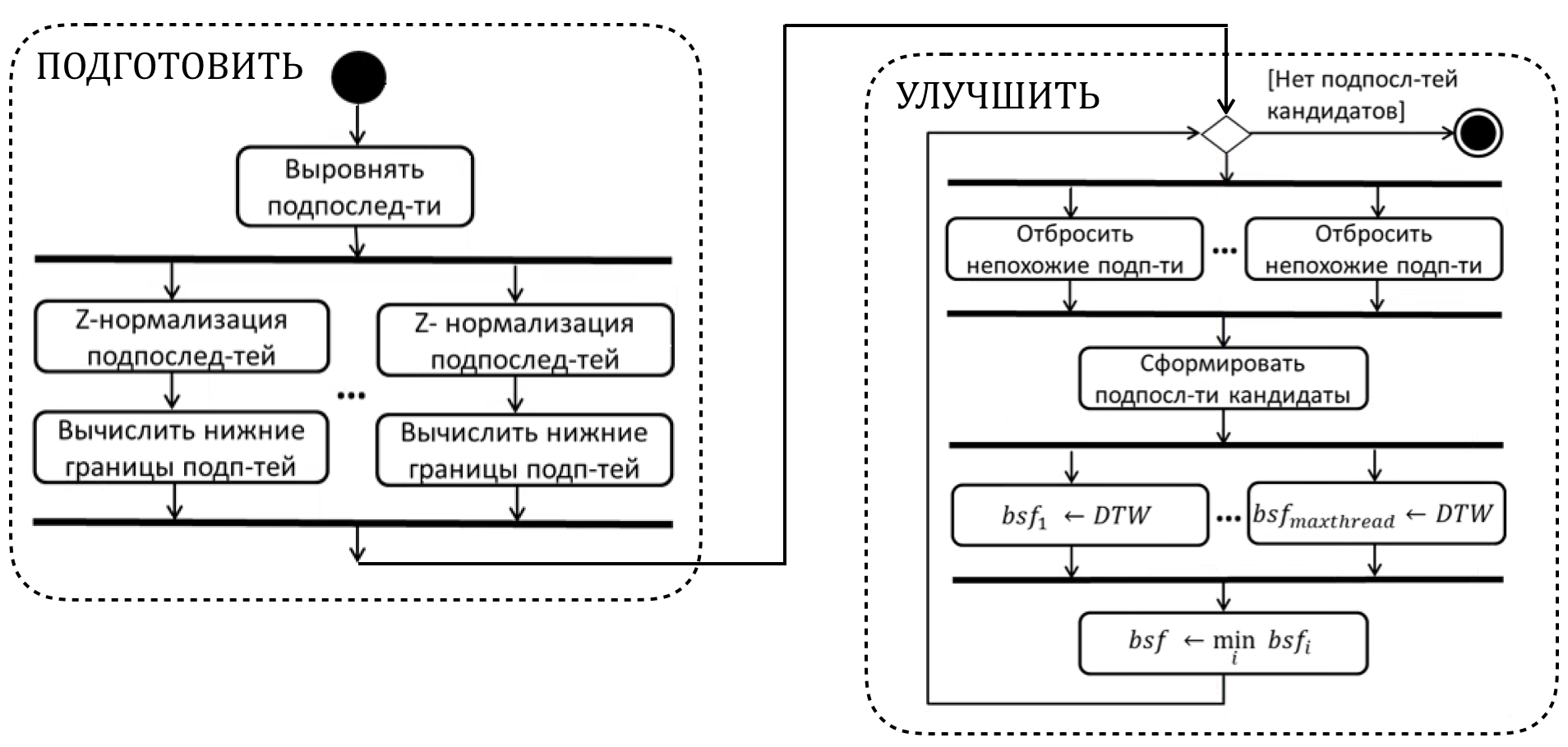}
	\caption{Схема вычислений алгоритма на одном узле кластера}
	\label{fig:ActionsOnNode}
\end{figure}

Предлагаемая в данной работе параллельная реализация поиска похожих подпоследовательностей временного ряда представлена в алг.~\ref{alg:PhiBestMatch} и на рис.~\ref{fig:ActionsOnNode}. 
Алгоритм выполняется следующим образом. 

\begin{algorithm}[!ht]
	\caption{\textsc{PhiBestMatch}(\textsc{in} $T, Q, r$; \textsc{out} $bsf, bestmatch$)}
	\label{alg:PhiBestMatch}
	\begin{algorithmic}[1]
		\State{$myrank \leftarrow~$\textsf{MPI\_Comm\_rank()}}
		\State{$N \leftarrow |T^{(myrank)}|-n+1$}
		\State{$subseq_{rnd} \leftarrow T^{(myrank)}_{random(1..N),\, n}$}		
		\State{$bsf \leftarrow $ \hyperref[alg:DTW]{\textsc{DTW}}$(subseq_{rnd}, Q, r, \infty)$}	
		\State{$processed \leftarrow N$}
		\State{\textsc{ПОДГОТОВИТЬ}}
		\Repeat
		\State{\textsc{УЛУЧШИТЬ}$(T^{(myrank)},\;bsf,\;bestmatch)$}
		\State{$flagDone \leftarrow (processed= 0)$}
		\State\parbox[t]{\dimexpr\linewidth-\algorithmicindent}{$\{bsf,bestmatch\} \leftarrow$ \textsf{MPI\_Allreduce(}$\{bsf,bestmatch\}$, \textsf{MPI\_FLOAT\_LONG}, \textsf{MPI\_MIN)}}			\State{$Stop \leftarrow$ \textsf{MPI\_Allreduce(}$myFragDone,\;$\textsf{MPI\_BOOL}, \textsf{MPI\_AND)}}
		\Until{\textbf{not} $Stop$}
		\State\Return{$\{ bsf, bestmatch \}$}
	\end{algorithmic}	
\end{algorithm}	

Для \emph{инициализации} вычислений алгоритм определяет номер текущего вычислительного процесса $myrank$ с помощью функций библиотеки MPI. В дальнейшем в рамках алгоритма каждый вычислительный процесс с номером $myrank$ обрабатывает матрицу подпоследовательностей $S^{n}_{T^{(myrank)}}$ фрагмента $T^{(myrank)}$ исходного временного ряда $T$. Переменная $bsf$ инициализируется значением меры схожести \emph{DTW} между поисковым запросом и случайной подпоследовательностью из данного фрагмента временного ряда. 

Далее выполняется \emph{подготовка данных} для вычислений. Формируется матрица выровненных подпоследовательностей. В цикле, выполняемом по строкам матрицы подпоследовательностей, каждая подпоследовательность z-нормализуется, и выполняются вычисление строк матрицы нижних границ схожести. Следует обратить внимание, что, строго говоря, предварительное вычисление матрицы нижних границ схожести представляет собой избыточные накладные расходы. В этом состоит ключевое отличие от алгоритма \emph{UCR-DTW}, где нижние границы схожести вычисляются каскадом (следующая нижняя граница вычисляется только в случае, если с помощью предыдущей границы  не выявлено, что текущий кандидат не является заведомо непохожим на запрос). Тем не менее, эти предвычисления обоснованы тем, что выполняются однократно, и могут быть реализованы с помощью распараллеливаемых и векторизуемых компилятором циклов.

Распараллеливание данного цикла осуществляется с помощью директивы OpenMP \texttt{\#pragma omp parallel for}, обеспечивающей статическое разбиение итераций цикла между нитями. Выровненные данные в матрице подпоследовательностей и отсутствие зависимостей по данным в вычислительных формулах нижних границ схожести обеспечивают векторизацию соответствующих вычислений.

После этого алгоритм выполняет следующий цикл действий, направленный на \emph{улучшение} значения порога \emph{bsf}. Цикл выполняется, пока каждый вычислительный узел не завершит обработку своего фрагмента. Сначала вычисляется карта схожести на основе предвычисленной матрицы нижних границ схожести. Полученное значение элемента карты схожести \texttt{FALSE} означает, что соответствующий элемент матрицы подпоследовательностей является заведомо непохожим на образец поиска и может быть отброшен без вычисления меры схожести \emph{DTW}. В противном случае этот элемент добавляется в матрицу кандидатов для последующего вычисления меры схожести \emph{DTW}. 

Для распараллеливания описанных выше операций матрица нижних границ схожести разбивается на сегменты (по числу используемых алгоритмом нитей). По построению матрица нижних границ схожести имеет существенно большее количество строк, чем матрица кандидатов. Соответственно, после заполнения и вычисления матрицы кандидатов, и обновления порога \emph{bsf} каждая нить должна продолжить сканирование своего сегмента вплоть до его исчерпания. Для хранения номера последнего обработанного кандидата в сегменте вводится индексный массив $Pos \in \mathbb{N}^p$, где 


\begin{equation}
\label{eq:SegIndex}
\begin{gathered}
pos_{i} :=  k: p \cdot (i-1)+1 \leqslant k \leqslant \lceil\tfrac{N}{i \cdot p}\rceil~\wedge \;
\forall j, 1 \leqslant j \leqslant lb_{max}, LB_{T}^n(k,j)<bsf.	
\end{gathered}
\end{equation} 

Отсутствие кандидатов означает, что обработка фрагмента закончена. В противном случае выполняется подсчет меры схожести \emph{DTW} для каждой строки матрицы кандидатов.

Чтобы вывести индекс самой похожей на запрос подпоследовательности, введен индексный массив $Idx \in \mathbb{N}^{s \cdot p}$, который предназначен для хранения позиции подпоследовательности во временном ряде и определяется следующим образом:

\begin{equation}
\label{eq:index}
idx_i := k: 1 \leqslant k \leqslant N \wedge \exists S_{T}^{n}(i,\cdot) \Leftrightarrow \exists T_{i,n} \Leftrightarrow k = (i-1) \cdot n+1.
\end{equation} 

После того, как матрица кандидатов заполнена, для каждой ее строки вычисляется значение меры схожести \emph{DTW} между запросом и кандидатом. Распараллеливание цикла осуществляется с помощью директивы OpenMP \texttt{\#pragma omp parallel for}, обеспечивающей статическое разбиение итераций цикла между нитями. Если вычисленное значение схожести меньше, чем текущее значение $bsf$, то $bsf$ заменяется на вычисленное значение. 

Выбор наилучшей среди всех фрагментов ряда нижней оценки схожести и соответствующей подпоследовательности осуществляется с помощью операции глобальной редукции \texttt{MPI\_Allreduce} стандарта MPI, которая возвращает минимальное значение порога \emph{bsf} среди всех вычислительных процессов и соответствующую подпоследовательность, копируя их в память каждого процесса.

Факт завершения обработки ряда также определяется с помощью операции глобальной редукции, в которой выполняется конъюнкция флагов завершения обработки каждым процессом своего фрагмента ряда.

\section{Вычислительные эксперименты.}
\label{sec:Experiments}

\subsection{Аппаратная платформа.}
\label{subsec:HardwareDatasetsGoals}

Для исследования эффективности разработанного алгоритма были проведены вычислительные эксперименты. В качестве аппаратной платформы экспериментов использовались узлы кластерной системы <<Торнадо ЮУрГУ>>~\cite{KostenetskyS16}, характеристики которых приведены в \tabref{tab:HardwarePhiBestMatch}.

\begin{table}[!ht]	
	\caption{Аппаратная платформа экспериментов}
	\label{tab:HardwarePhiBestMatch}
	\centering
	\begin{tabular}{|l|c|c|}
		\hline
		Характеристика & Хост & Сопроцессор \\ \hline
		Модель, Intel Xeon &  X5680 & Phi (KNC), SE10X \\ \hline
		Количество физических ядер & 2$\times$6 & 61 \\ \hline
		Гиперпоточность & 2$\times$ & 4$\times$ \\ \hline
		Количество логических ядер & 24 & 244 \\ \hline
		Частота, ГГц & 3.33 & 1.1 \\ \hline
		Размер VPU, бит & 128 & 512 \\ \hline
		Пиковая производительность, TFLOPS & 0.371 & 1.076 \\ \hline
	\end{tabular}
\end{table}

В экспериментах исследована эффективность алгоритма \emph{PhiBestMatch} как на одном вычислительном узле кластерной системы, так и на кластерной системе в целом. 

Исследование на одном вычислительном узле кластерной системы позволяет определить, насколько эффективно реализованы параллельные вычисления, выполняемые многоядерной процессорной системой Intel Xeon Phi. Для такого исследования использовалась упрощенная версия алг.~\ref{alg:PhiBestMatch}~\cite{KraevaZ18}, в которой ряд отождествлен с одним фрагментом, вызовы коммуникационной библиотеки MPI не задействуются, и цикл обработки подпоследовательностей ряда продолжается до исчерпания подпоследовательностей-кандидатов. 

Во всех экспериментах размер сегмента матрицы кандидатов~--- количество строк матрицы кандидатов, обрабатываемых одной нитью в рамках одного вычислительного узла кластерной системы (см. формулу~\ref{eq:candidateMatrix})~--- имеет значение $s=100$.

\subsection{Эффективность алгоритма на одном узле кластера.}

Исследование эффективности алгоритма на одном вычислительном узле кластера <<Торнадо ЮУрГУ>> производилось на наборах данных, которые представлены в \tabref{tab:DatasetsSingleNodePhiBestMatch}.

\begin{table}[!ht]	
	\caption{Наборы данных для экспериментов на одном узле кластера}
	\label{tab:DatasetsSingleNodePhiBestMatch}
	\centering
	\begin{tabular}{|l|c|c|c|}
		\hline
		Набор данных & Вид & $|T|=m$ & $|Q|=n$ \\ \hline
		Random Walk &  Синтетический & $10^6$ & 128 \\ \hline
		EPG & Реальный & $2.5\cdot 10^5$ & 360 \\ \hline
	\end{tabular}
\end{table}

Временной ряд Random Walk получен искусственно на основе модели случайных блужданий~\cite{Pearson1905}. Проведение экспериментов по исследованию свойств алгоритмов поиска подпоследовательностей на базе меры схожести \emph{DTW} с использованием подобных временных рядов является общепринятой практикой (см., например, работы~\cite{sart,DBLP:conf/icacci/ShabibNNDPSATS15,wang}). 

Временной ряд EPG (Electrical Penetration Graph, график электрического проникновения) использован в экспериментах в работе~\cite{sart}. EPG представляет собой набор сигналов, используемых энтомологами для сравнения поведения исследуемого насекомого с поведением цикадок \emph{macrosteles quadrilineatus}, которые являются переносчиками болезней растений и ежегодно наносят сельскому хозяйству США ущерб более чем на 2~млн. долларов~\cite{sart}.

В экспериментах исследовались производительность и масштабируемость алгоритма \emph{PhiBestMatch}. Под производительностью понимается время работы алгоритма без учета времени загрузки данных в память и выдачи результата. Масштабируемость параллельного алгоритма означает его способность адекватно адаптироваться к увеличению параллельно работающих вычислительных элементов (процессов, процессоров, нитей и др.) и характеризуется ускорением и параллельной эффективностью, которые определяются следующим образом~\cite{VoevodinV02}. \emph{Ускорение} и \emph{параллельная эффективность} параллельного алгоритма, запускаемого на $k$~нитях, вычисляются как $s(k)=\tfrac{t_1}{t_k}$ и $e(k)=\tfrac{s(k)}{k}$
соответственно, где $t_1$ и $t_k$~--- время работы алгоритма на одной и $k$
нитях соответственно.

В экспериментах рассматривались вышеупомянутые показатели в зависимости от изменения параметра $r$ (ширина полосы Сако---Чиба), значения которого брались в долях длины поискового запроса~$n$. 

Время работы алгоритма \emph{PhiBestMatch} на многоядерных платформах сравнивалось с временем работы  алгоритма \emph{UCR-DTW}~\cite{rakthanmanon}.

\begin{figure}[!ht]
	\begin{minipage}[t]{0.5\textwidth}
		\includegraphics[width=\linewidth]{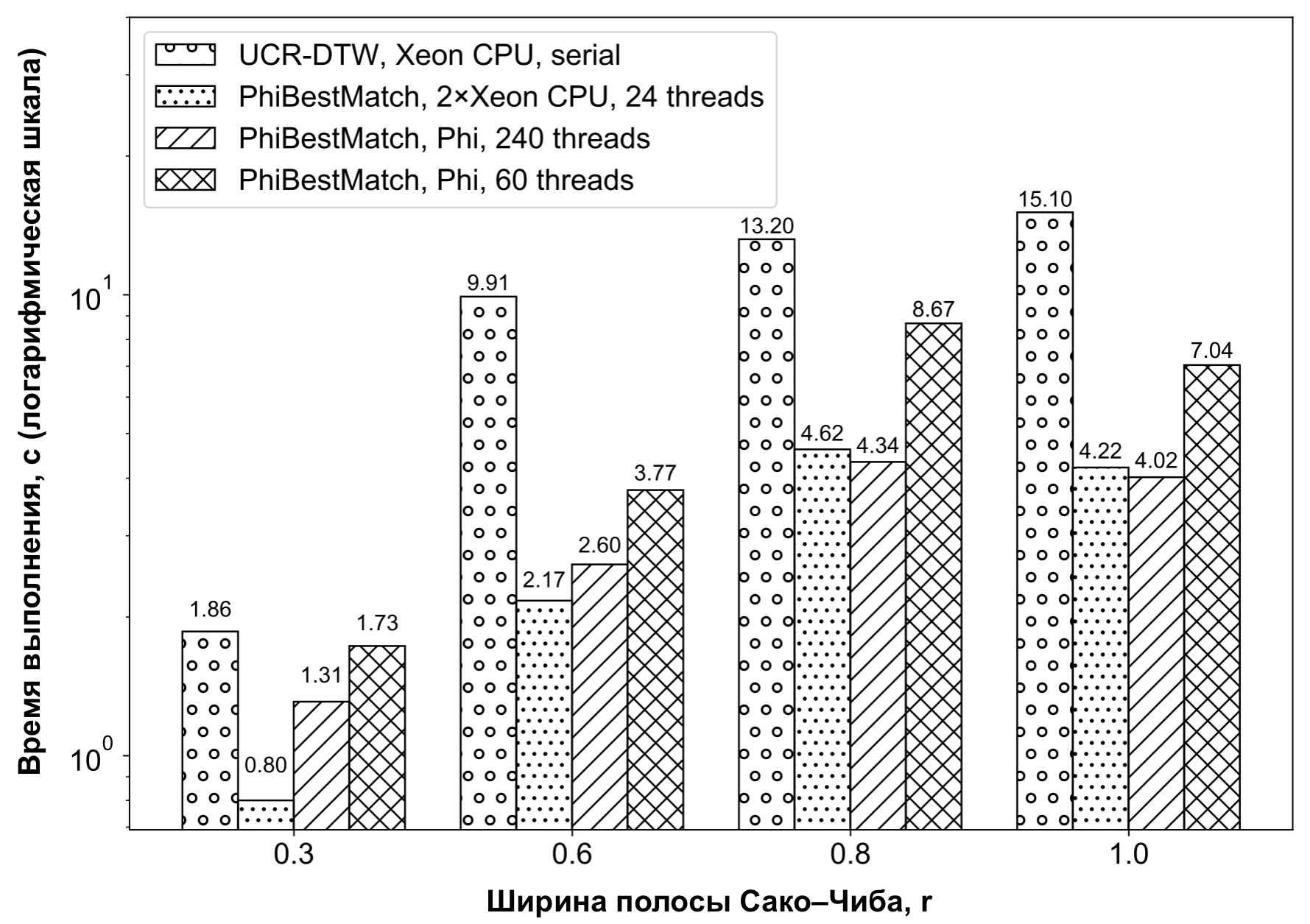}
		\center{а) Набор синтетических данных Random Walk, $m=10^6, n=128$}
		\label{subfig:speedup-RandomWalk}
	\end{minipage}
	\hfill
	\begin{minipage}[t]{0.5\textwidth}
		\includegraphics[width=\linewidth]{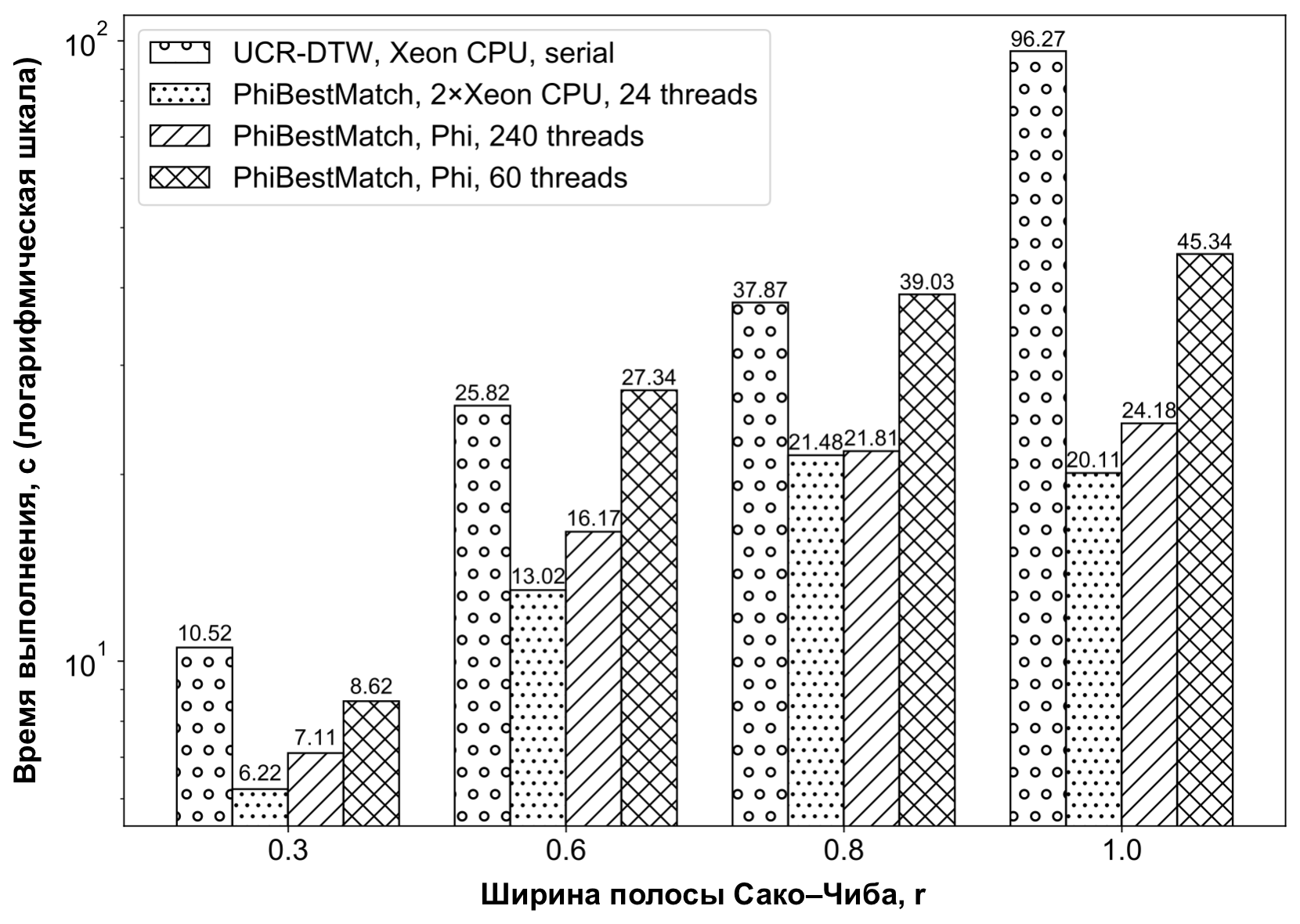}		
		\center{б) Набор реальных данных EPG, $m=2.5\cdot10^5, n=360$}
		\label{subfig:efficiency-RandomWalk}
	\end{minipage}
	\hspace{1em}
	\caption{Производительность алгоритма \emph{PhiBestMatch} на одном узле кластера }
	\label{fig:PhiBestMatch-SingleNode-Runtime}
\end{figure}

Результаты экспериментов по исследованию производительности алгоритма представлены на рис.~\ref{fig:PhiBestMatch-SingleNode-Runtime}. Можно видеть, что \emph{PhiBestMatch} работает до 5~раз быстрее, чем алгоритм \emph{UCR-DTW}. Вместе с тем видно, на производительность алгоритма \emph{PhiBestMatch} на платформах двухпроцессорного узла Intel Xeon и многоядерной системы Intel Xeon Phi влияют следующие два параметра: ширина полосы Сако---Чиба $r$ и длина поискового запроса $n$.

При малых значениях этих параметров (примерно $0 < r \leqslant 0.5n$ и $n<512$) алгоритм \emph{PhiBestMatch} на платформе двухпроцессорного узла Intel Xeon работает несколько быстрее или примерно с тем же быстродействием, что и на платформе многоядерной системы Intel Xeon Phi. При б\'{о}льших значениях данных параметров ($0.5n<r \leqslant n$ и $n \geqslant 512$) алгоритм работает быстрее на платформе Intel Xeon Phi. Это означает, что заложенные в алгоритм при проектировании возможности векторизации наилучшим образом проявляются при увеличении общего объема вычислений. 

Поисковые запросы с длиной $n \geqslant 512$, равно как и значение параметра $r=1$ требуются на практике в ряде приложений, требующих при определении схожести подпоследовательностей как можно более высокую точность, например, в медицине при исследовании ЭКГ~\cite{DBLP:journals/cmpb/MazandaraniM18}, в 
энтомологии~\cite{DBLP:conf/sigmod/AthitsosPPKG08}, в астрономии~\cite{Rebbapragada2009} и др.  

\begin{figure}[!ht]
	\begin{minipage}[h]{0.5\textwidth}
		\includegraphics[width=\linewidth]{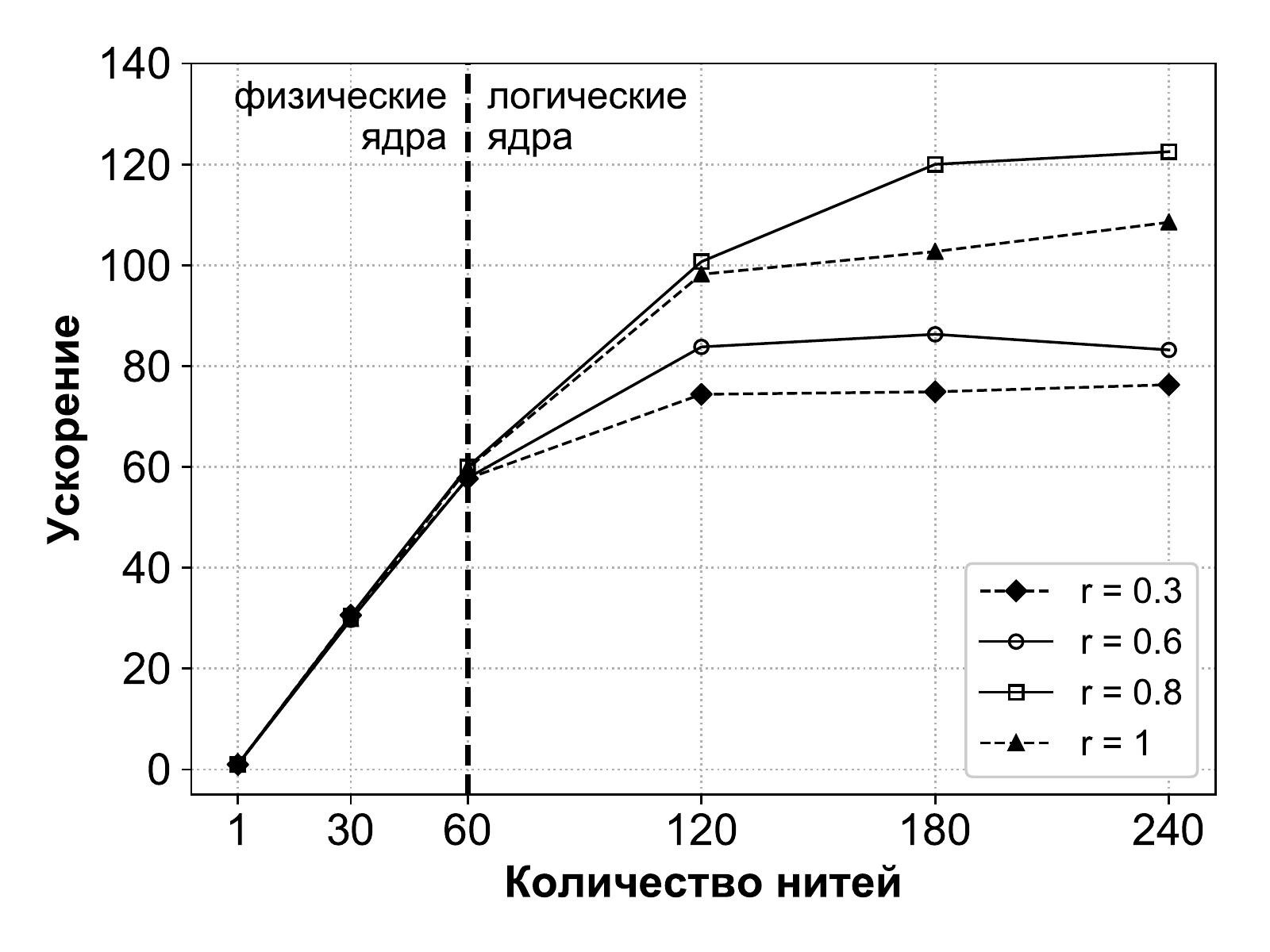}
		\center{а) Ускорение}
		\label{subfig:speedup-RandomWalk}
	\end{minipage}
	\hfill
	\begin{minipage}[h]{0.5\textwidth}
		\includegraphics[width=\linewidth]{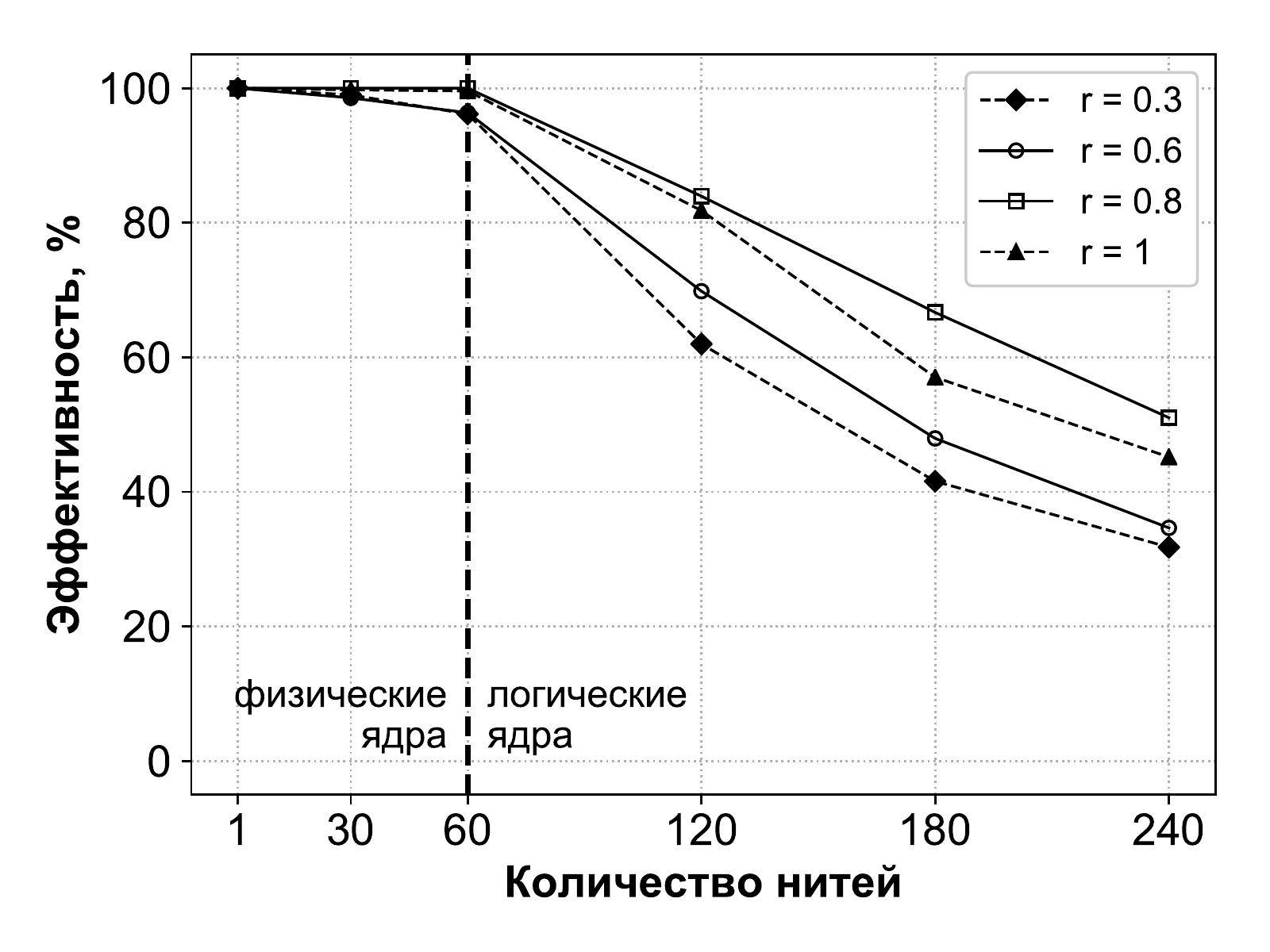}		
		\center{б) Параллельная эффективность}
		\label{subfig:efficiency-RandomWalk}
	\end{minipage}
	\hspace{1em}
	\caption{Масштабируемость алгоритма \emph{PhiBestMatch} на одном узле кластера при обработке синтетических данных (ряд Random Walk, $m=10^6, n=128$)}
	\label{fig:PhiBestMatch-SingleNode-Scalability-RandomWalk}
\end{figure}

\begin{figure}[!ht]
	\begin{minipage}[h]{0.5\textwidth}
		\includegraphics[width=\linewidth]{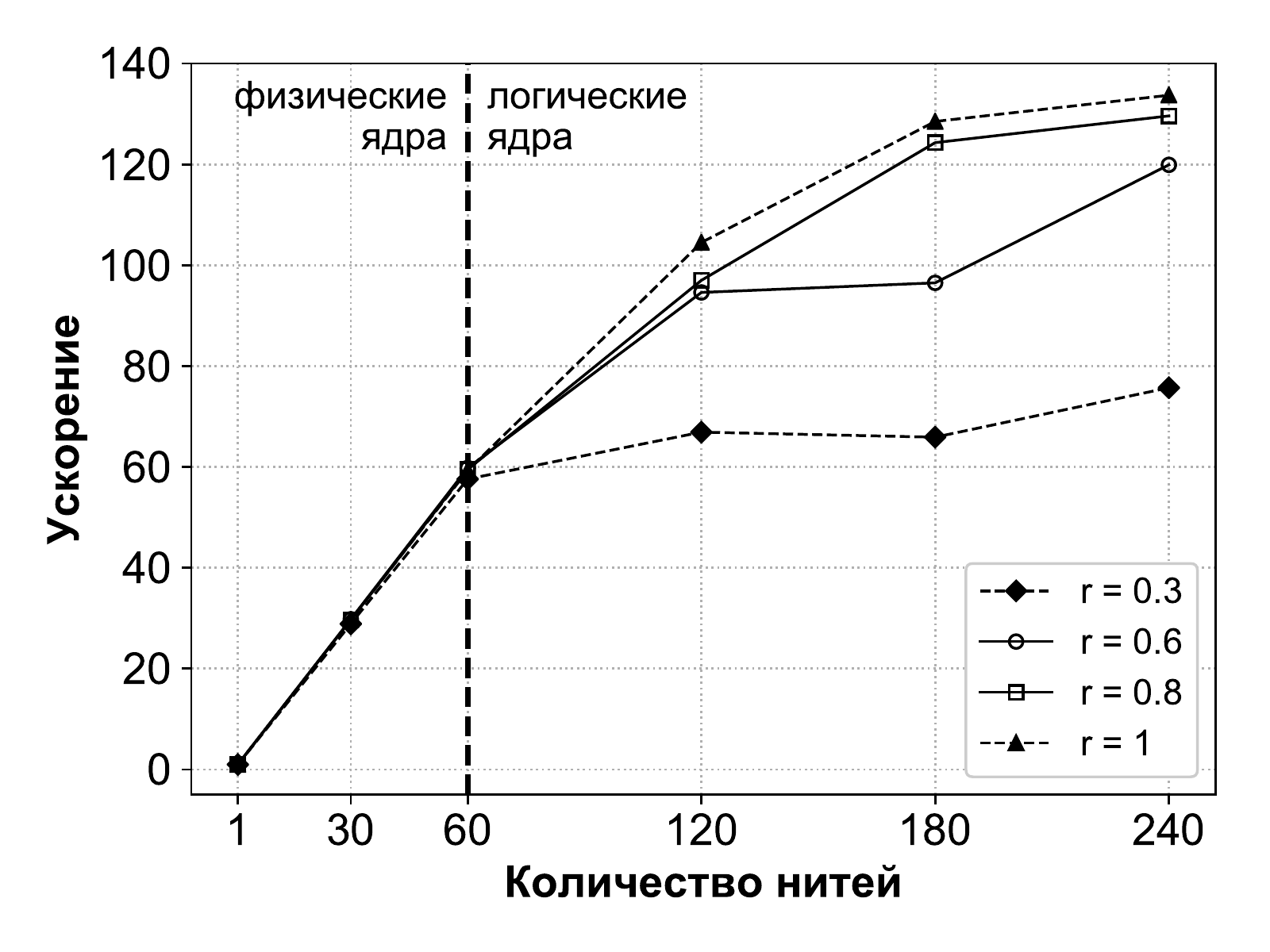}
		\center{а) Ускорение}
		\label{subfig:speedup-EPG}
	\end{minipage}
	\hfill
	\begin{minipage}[h]{0.5\textwidth}
		\includegraphics[width=\linewidth]{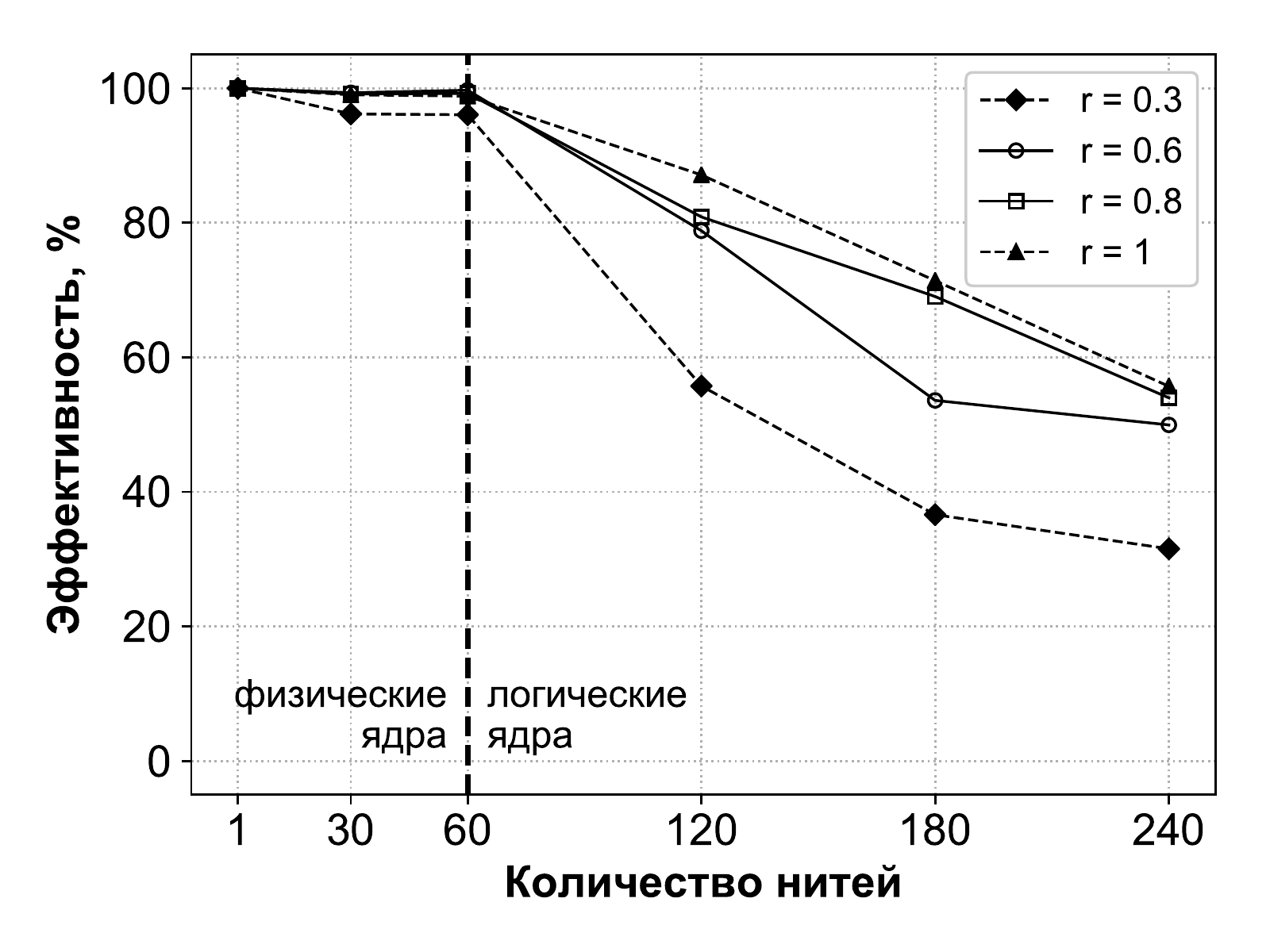}		
		\center{б) Параллельная эффективность}
		\label{subfig:efficiency-EPG}
	\end{minipage}
	\hspace{1em}
	\caption{Масштабируемость алгоритма \emph{PhiBestMatch} на одном узле кластера при обработке реальных данных (ряд EPG, $m=2.5\cdot10^5, n=360$)}
	\label{fig:PhiBestMatch-SingleNode-Scalability-EPG}
\end{figure}

Результаты экспериментов по исследованию масштабируемости алгоритма представлены на \figref{fig:PhiBestMatch-SingleNode-Scalability-RandomWalk}~и~\ref{fig:PhiBestMatch-SingleNode-Scalability-EPG}. Можно видеть, что \emph{PhiBestMatch} демонстрирует близкое к линейному ускорение и параллельную эффективность, близкую к 100\%, если количество нитей, на которых запущен алгоритм, совпадает с количеством физических ядер системы Intel Xeon Phi.

При увеличении количества нитей, запускаемых на одном физическом ядре системы, ускорение становится сублинейным, равно как наблюдается и падение параллельной эффективности. При этом наилучшие показатели ускорения и параллельной эффективности ожидаемо наблюдаются при значениях параметра $r$ 0.8 и 1 от длины поискового запроса $n$, обеспечивающих алгоритму наибольшую вычислительную нагрузку. Например, при задействовании 240~нитей при $r=0.8n$ обработка ряда Random Walk выполняется с ускорением~120 и параллельной эффективностью 50\%; при $r=n$ обработка ряда EPG~--- с ускорением~130 и параллельной эффективностью 52\%. 

Полученные результаты позволяют сделать заключение о хорошей масштабируемости разработанного алгоритма и эффективном использовании им возможностей векторизации вычислений на многоядерной системе Intel Xeon Phi в рамках одного вычислительного узла кластерной системы. Указанные свойства проявляются при значениях параметров $r$ (ширина полосы Сако---Чиба) и $n$ (длина поискового запроса), обеспечивающих алгоритму наибольшую вычислительную нагрузку: $0.8n \leqslant r \leqslant n$ и $n \geqslant 512$ соответственно. 

\subsection{Эффективность алгоритма на кластерной системе.}
\label{subsubsec:chap-TimeSeriesMining-sec-PhiBestMatch-subsec-Experiments-Cluster}

В исследовании эффективности алгоритма на кластерной системе в целом было задействовано от 16 до 128~вычислительных узлов суперкомпьютера <<Торнадо ЮУрГУ>> (см. \tabref{tab:HardwarePhiBestMatch}). Исследование производилось на наборах данных, которые представлены в \tabref{tab:DatasetsAllNodesPhiBestMatch}.

\begin{table}[!ht]	
	\caption{Наборы данных для экспериментов на кластерной системе}
	\label{tab:DatasetsAllNodesPhiBestMatch}
	\centering
	\begin{tabular}{|l|c|c|c|}
		\hline
		Набор данных & Вид & $|T|=m$ & $|Q|=n$ \\ \hline
		Random Walk &  Синтетический & $12.8 \cdot 10^7$ & 128, 512, 1024 \\ \hline
		ECG & Реальный & $12.8 \cdot 10^7$ & 432, 512, 1024 \\ \hline
	\end{tabular}
\end{table}

В экспериментах исследовалось \emph{ускорение масштабируемости (scaled speedup)} параллельного алгоритма, которое определяется как ускорение, демонстрируемое алгоритмом при линейном увеличении объема данных и количества используемых вычислительных узлов~\cite{DBLP:books/bc/KumarGGK94} и вычисляется следующим образом: $s_{scaled}=\tfrac{p \cdot m}{t_{p(p \cdot m)}}$, где $p$~--- количество задействованных вычислительных узлов, $m$~--- объем исходных данных, $t_{p(p \cdot m)}$~--- время выполнения алгоритма на $p$ узлах при обработке исходных данных, имеющих объем $p \cdot m$. 

При этом значение параметра $r$ было зафиксировано как $r=n$ и варьировалась длина поискового запроса.

\begin{figure}[!ht]
	\center	
	\includegraphics[width=1\linewidth]{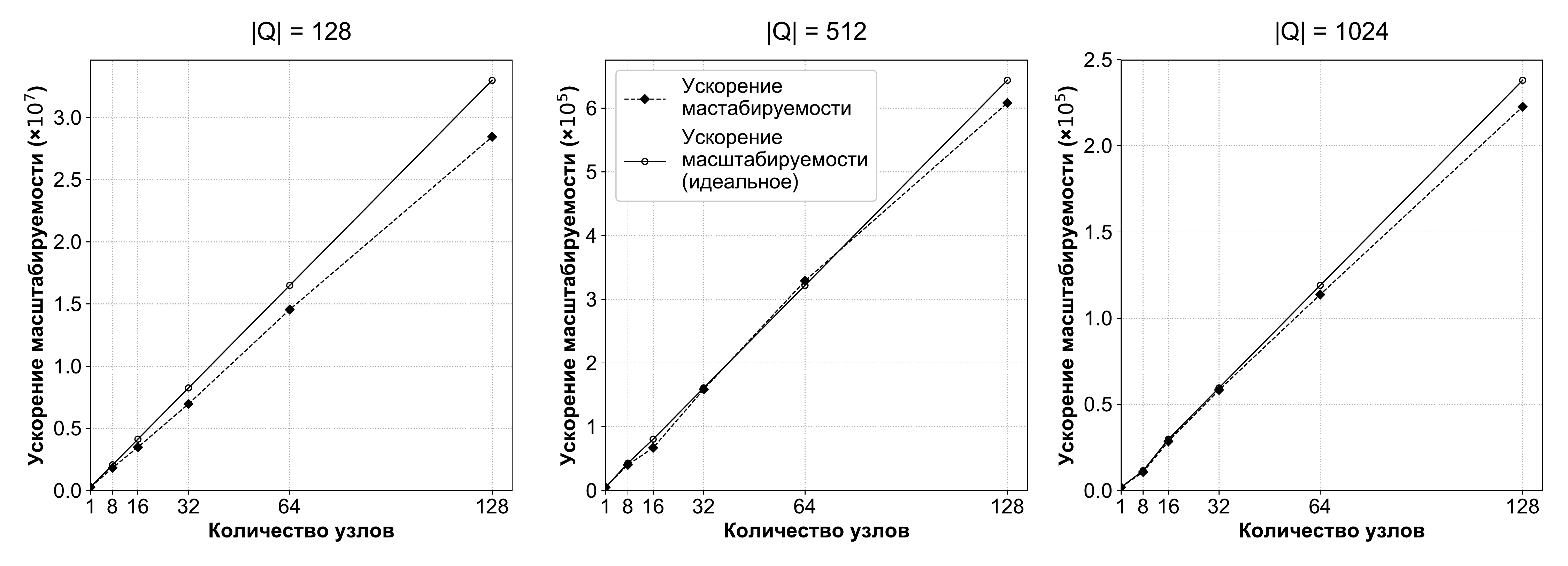}
	\caption{Ускорение масштабируемости алгоритма \emph{PhiBestMatch} на кластерной системе при обработке синтетических данных}
	\label{fig:PhiBestMatch-AlleNodes-ScaledSpeedup-RandomWalk}
\end{figure}

\begin{figure}[!ht]
	\center
	\includegraphics[width=1\linewidth]{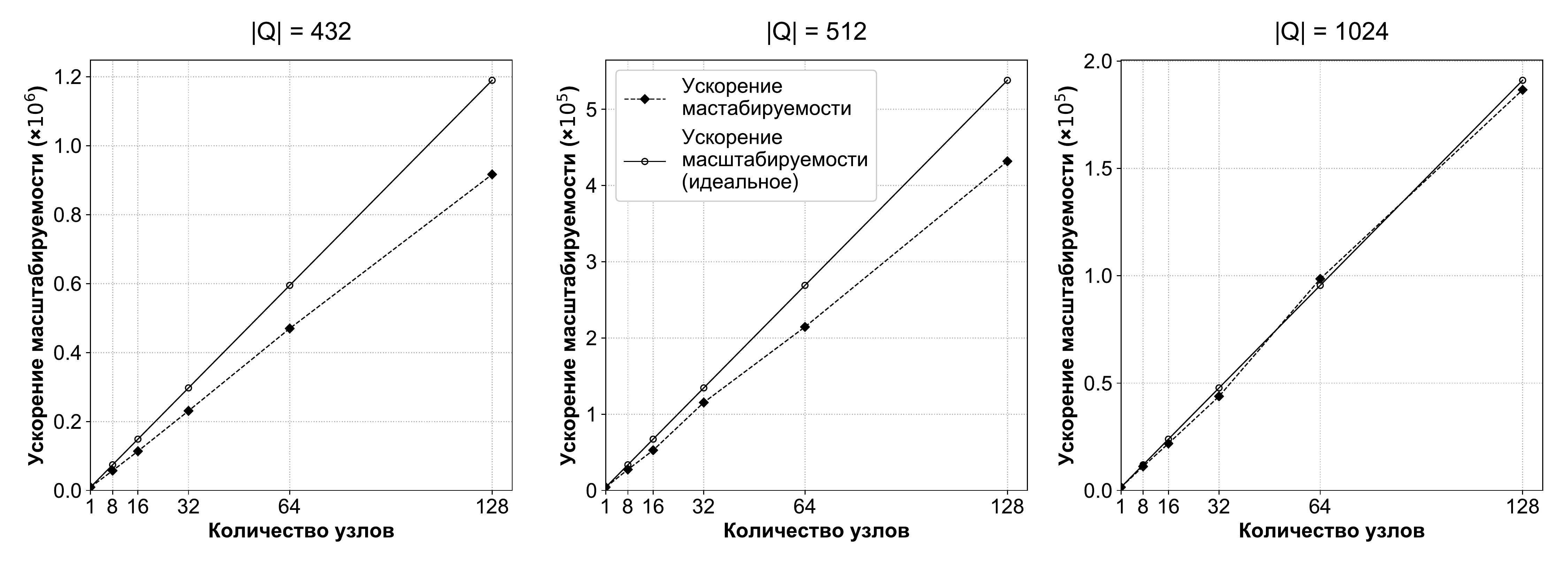}
	\caption{Ускорение масштабируемости алгоритма \emph{PhiBestMatch} на кластерной системе при обработке реальных данных}
	\label{fig:PhiBestMatch-AlleNodes-ScaledSpeedup-EPG}
\end{figure}

Результаты экспериментов по исследованию ускорения масштабируемости алгоритма представлены на \figref{fig:PhiBestMatch-AlleNodes-ScaledSpeedup-RandomWalk} и \ref{fig:PhiBestMatch-AlleNodes-ScaledSpeedup-EPG}.

Можно видеть, что \emph{PhiBestMatch} демонстрирует  ускорение масштабируемости, близкое к линейному, как для синтетических, так и для реальных данных. При этом поиск подпоследовательности большей длины показывает более высокое ускорение масштабируемости, поскольку это обеспечивает больший объем вычислений в рамках одного вычислительного узла кластерной системы.

Полученные результаты позволяют сделать заключение о хорошей масштабируемости разработанного алгоритма при работе на кластерной системе с вычислительными узлами на базе многоядерных процессоров Intel Xeon Phi. Указанные свойства проявляются при значениях параметров $r$ (ширина полосы Сако---Чиба) и $n$ (длина поискового запроса), обеспечивающих алгоритму наибольшую вычислительную нагрузку: $0.8n \leqslant r \leqslant n$ и $n \geqslant 512$ соответственно.

\section{Заключение.}
\label{sec:Conclusion}
В работе рассмотрена проблема поиска похожих подпоследовательностей в сверхбольших временных рядах (сотни миллиардов точек) на основе использования меры схожести \emph{DTW} (динамическая трансформация шкалы времени). Данная задача возникает в широком спектре приложений интеллектуального анализа временных рядов: моделирование климата, финансовые прогнозы, медицинские исследования и др.

Предложен новый параллельный алгоритм для поиска похожих подпоследовательностей временного ряда на кластерной системе с вычислительными узлами на базе многоядерных процессоров Intel Xeon Phi поколения Knights Landing, названный \emph{PhiBestMatch}. Алгоритм предполагает двухуровневое распараллеливание вычислений: на уровне всех узлов кластерной системы используется технология MPI, в рамках одного вычислительного узла кластера используется технология OpenMP. Для эффективного использования возможностей векторизации вычислений процессоров Phi KNL в рамках одного вычислительного узла используются дополнительные матричные структуры данных и избыточные вычисления. Проведены вычислительные эксперименты, исследующие быстродействие и масштабируемость алгоритма на синтетических и реальных наборах данных как на кластерной системе в целом, так и в рамках одного вычислительного узла кластера. Результаты  экспериментов показали хорошую масштабируемость алгоритма \emph{PhiBestMatch} на одном узле и на кластерной системе в целом.

\vspace{1em}
{Работа выполнена при финансовой поддержке Российского фонда фундаментальных исследований (грант №~17-07-00463), Правительства РФ в соответствии с Постановлением №~211 от 16.03.2013 (соглашение №~02.A03.21.0011) и Министерства образования и науки РФ (государственное задание 2.7905.2017/8.9).}

\begin{biblio}
	
	
	\bibitem{VoevodinV02}
	\emph{Воеводин~В.В., Воеводин~Вл.В.} Параллельные вычисления. СПб.:
	БХВ-Петербург, 2002.
	
	
	
	\bibitem{DBLP:conf/sigmod/AthitsosPPKG08}
	\emph{Athitsos~V., Papapetrou~P., Potamias~M. et al.} Approximate embedding-based subsequence matching of time series // Proc. of the ACM SIGMOD Int. Conf. on Management of Data (Vancouver, Canada, June 10--12, 2008). New York: ACM, 2008. 365--378.  DOI:~\href{https://dx.doi.org/10.1145/1376616.1376656}{10.1145/1376616.1376656}.
	
	\bibitem{bacon}
	\emph{Bacon~D.F., Graham~S.L., Sharp~O.J.} Compiler transformations for high-performance computing // ACM Comput. Surv. 1994. \textbf{26}, N~4. 345--420. DOI:~\href{https://dx.doi.org/10.1145/197405.197406}{10.1145/197405.197406}.
	
	\bibitem{DBLP:conf/kdd/BerndtC94}
	\emph{Berndt~D.J., Clifford~J.} Using dynamic time warping to find patterns in time series // Proc. of the 1994 AAAI Workshop on Knowledge Discovery in Databases (Seattle, Washington, July, 1994). AAAI Press, 1994. 359--370.
	
	\bibitem{chrysos}
	\emph{Chrysos~G.} Intel Xeon Phi Coprocessor (codename Knights Corner) // 2012 IEEE Hot Chips 24th Symposium (HCS) (Cupertino, CA, USA, August~27--29, 2012). IEEE, 2012. 1--31. DOI:~\href{https://dx.doi.org/10.1109/HOTCHIPS.2012.7476487}{10.1109/HOTCHIPS.2012.7476487}.
	
	\bibitem{DBLP:journals/pvldb/DingTSWK08}
	\emph{Ding~H., Trajcevski~G., Scheuermann~P., Wang~X., Keogh~E.} Querying and mining of time series data: experimental comparison of representations and distance measures // Proc. of the VLDB Endowment. 2008. \textbf{1}, N~2. 1542--1552.
	
	\bibitem{fu}
	\emph{Fu~A.W., Keogh~E.J., Lau~L.Y.H., Ratanamahatana~C.A., Wong~R.C.} Scaling and time warping in time series querying // VLDB. 2008. \textbf{17}, N~4. 899--921. DOI:~\href{https://doi.org/10.1007/s00778-006-0040-z}{10.1007/s00778-006-0040-z}.
	
	\bibitem{DBLP:conf/pvm/Gropp12}
	\emph{Gropp W.} MPI 3 and beyond: why MPI is successful and what challenges it faces // Proc. of the 19th European {MPI} Users' Group Meeting. Lecture Notes in Computer Science. Vol.~7490. Berlin: Springer, 2012. 1--9. DOI:~\href{https://dx.doi.org/10.1007/978-3-642-33518-1\_1}{10.1007/978-3-642-33518-1\_1}.
	
	\bibitem{keogh}
	\emph{Keogh~E.J., Ratanamahatana~C.A.} Exact indexing of dynamic time warping // Knowl. Inf. Syst. 2005. \textbf{7}, N~3. 358--386. DOI:~\href{https://doi.org/10.1007/s10115-004-0154-9}{10.1007/s10115-004-0154-9}.
	
	
	\bibitem{KostenetskyS16}
	\emph{Kostenetskiy~P.S., Safonov~A.Y.} SUSU supercomputer resources // Proc. of the 10th Annual Int. Scientific Conf. on Parallel Computing Technologies (PCT 2016). CEUR Workshop Proceedings. Vol.~1576. 2016. 561--573.
	
	\bibitem{KraevaZ18}
	\textit{Kraeva~Ya., Zymbler~M.} An Efficient Subsequence Similarity Search on Modern Intel Many-core Processors for Data Intensive Applications // Аналитика и управление данными в областях с интенсивным использованием данных: Сборник научных трудов XX Международной конференции DAMDID / RCDL'2018 (9--12 октября 2018 г., Москва, Россия). С.~116--124.
	
	\bibitem{DBLP:books/bc/KumarGGK94}
	\emph{Kumar~V., Grama~A., Gupta~A., Karypis~G.} Introduction to parallel
	computing. Benjamin/Cummings, 1994.
	
	\bibitem{lim}
	\emph{Lim~S., Park~H., Kim~S.} Using multiple indexes for efficient subsequence matching in time-series databases // Proc. of the 11th Int. Conf. on Database Systems for Advanced Applications. Lecture Notes in Computer Science. Vol.~3882. Berlin: Springer, 2006. 65--79. DOI:~\href{https://doi.org/10.1007/11733836\_7}{10.1007/11733836\_7}.
	
	\bibitem{DBLP:conf/sc/Mattson06}
	\emph{Mattson~T.} Introduction to OpenMP // Proc. of the 2006 ACM/IEEE Conf. on Supercomputing (Tampa, FL, USA, November~11--17, 2006). New York: ACM Press, 2006. 209.
	DOI:~\href{https://dx.doi.org/10.1145/1188455.1188673}{10.1145/1188455.1188673}.
	
	\bibitem{DBLP:journals/cmpb/MazandaraniM18}
	\emph{Mazandarani~F.N., Mohebbi~M.} Wide complex tachycardia discrimination using dynamic time warping of ECG beats // Computer Methods and Programs in Biomedicine. 2018. \textbf{164}. 238--249. DOI:~\href{https://dx.doi.org/10.1016/j.cmpb.2018.04.009}{10.1016/j.cmpb.2018.04.009}.
	
	\bibitem{MovchanZ16}
	\emph{Movchan~A.V., Zymbler~M.L.} Parallel implementation of searching the most similar subsequence in time series for computer systems with distributed memory // Proc. of the 10th Annual Int. Scientific Conf. on Parallel Computing Technologies (PCT 2016). CEUR Workshop Proceedings. Vol.~1576. 2016. 615--628. 
	
	\bibitem{Pearson1905}
	\emph{Pearson~K.} The problem of the random walk // Nature. 1905. \textbf{72}, N~1865. 294. DOI:~\href{https://dx.doi.org/10.1038/072342a0}{10.1038/072342a0}.
	
	\bibitem{rakthanmanon}
	\emph{Rakthanmanon~T., Campana~B.J.L., Mueen~A., Batista~G.E.A.P.A., Westover~M.B., et al.} Searching and mining trillions of time series subsequences under dynamic time warping // Proc. of the 18th ACM SIGKDD Int. Conf. on Knowledge Discovery and Data Mining (Beijing, China, August~12--16, 2012). New York: ACM, 2012. 262--270. DOI:~\href{http://doi.acm.org/10.1145/2339530.2339576}{10.1145/2339530.2339576}.
	
	\bibitem{Rebbapragada2009}	
	\emph{Rebbapragada~U., Protopapas~P., Brodley~C.E., Alcock~C.} Finding anomalous periodic time series // Machine Learning. 2009. \textbf{74}, N~3. 281--313. DOI:~\href{https://dx.doi.org/10.1007/s10994-008-5093-3}{10.1007/s10994-008-5093-3}.	
	
	\bibitem{sakoe}
	\emph{Sakoe~H., Chiba~S.} Readings in speech recognition. San Francisco, CA, USA: Morgan Kaufmann Publishers Inc., 1990. 159--165.
	
	\bibitem{sakurai}
	\emph{Sakurai~Y., Faloutsos~C., Yamamuro M.} Stream monitoring under the time warping distance // Proc. of the 23rd Int. Conf. on Data Engineering (Istanbul, Turkey, April~15--20, 2007). Washington, DC, USA: IEEE Computer Society, 2007. 1046--1055. DOI:~\href{https://dx.doi.org/10.1109/ICDE.2007.368963}{10.1109/ICDE.2007.368963}.	
	
	\bibitem{sart}
	\emph{Sart~D., Mueen~A., Najjar~W.A., Keogh~E.J., Niennattrakul~V.} Accelerating dynamic time warping subsequence search with GPUs and FPGAs // Proc. of the 2010 IEEE Int. Conf. on Data Mining (Sydney, Australia, December~14--17, 2010). Washington, DC, USA: IEEE Computer Society, 2010. 1001--1006. DOI:\href{https://doi.org/10.1109/ICDM.2010.21}{10.1109/ICDM.2010.21}.
	
	\bibitem{sodani}
	\emph{Sodani~A.} Knights Landing (KNL): 2nd generation Intel Xeon Phi processor // 2015 IEEE Hot Chips 27th Symposium (HCS) (Cupertino, CA, USA, August~22--25, 2015). IEEE, 2015. 1--24.	
	
	\bibitem{SokolinskayaS16}
	\emph{Sokolinskaya~I.,  Sokolinsky~L.} Revised pursuit algorithm for solving non-stationary linear programming problems on modern computing clusters with manycore accelerators // Communications in Computer and Information Science. 2016. \textbf{687}. 212--223. DOI:~\href{https://dx.doi.org/10.1007/978-3-319-55669-7\_17}{10.1007/978-3-319-55669-7\_17}.
	
	\bibitem{DBLP:conf/icacci/ShabibNNDPSATS15}
	\emph{Shabib~A., Narang~A., Niddodi~C.P. et al.} Parallelization of searching and mining time series data using Dynamic Time Warping // 2015 Int. Conf. on Advances in Computing, Communications and Informatics (Kochi, India, August 10--13, 2015). IEEE, 2015. 343–348. DOI:~\href{https://dx.doi.org/10.1109/ICACCI.2015.7275633}{10.1109/ICACCI.2015.7275633}.
	
	\bibitem{srikanthan}
	\emph{Srikanthan~S., Kumar~A., Gupta~R.} Implementing the dynamic time warping algorithm in multithreaded environments for real time and unsupervised pattern discovery // 2011 2nd Int. Conf. on Computer and Communication Technology (Allahabad, India, September~15--17, 2011). IEEE, 2015. 394--398. DOI:~\href{https://doi.org/10.1109/ICCCT.2011.6075111}{10.1109/ICCCT.2011.6075111}.
	
	\bibitem{takahashi}
	\emph{Takahashi~N., Yoshihisa~T., Sakurai~Y., Kanazawa~M.} A parallelized data stream processing system using dynamic time warping distance // 2009 Int. Conf. on Complex, Intelligent and Software Intensive Systems (Fukuoka, Japan, March~16--19, 2009). IEEE, 2009. 1100--1105. DOI:~\href{https://doi.org/10.1109/CISIS.2009.77}{10.1109/CISIS.2009.77}.
	
	\bibitem{tarango}
	\emph{Tarango~J., Keogh~E.J., Brisk~P.} Instruction set extensions for dynamic time warping // Proc. of the Int. Conf. on Hardware/Software Codesign and System Synthesis (Montreal, QC, Canada, September~29--October~4, 2013). IEEE, 2013. 18:1--18:10. DOI:~\href{https://doi.org/10.1109/CODES-ISSS.2013.6659005}{10.1109/CODES-ISSS.2013.6659005}.
	
	\bibitem{wang}
	\emph{Wang~Z., Huang~S., Wang~L., Li~H., Wang~Y., et al.} Accelerating subsequence similarity search based on dynamic time warping distance with FPGA // Proc. of the ACM/SIGDA Int. Symposium on Field Programmable Gate Arrays (Monterey, CA, USA, February~11--13, 2013). New York: ACM, 2013. 53--62. DOI:~\href{http://doi.acm.org/10.1145/2435264.2435277}{10.1145/2435264.2435277}.
	
	\bibitem{zhang}
	\emph{Zhang~Y., Adl~K., Glass~J.R.} Fast spoken query detection using
	lower-bound dynamic time warping on graphical processing units // 2012 IEEE Int. Conf. on Acoustics, Speech and Signal Proc. (Kyoto, Japan, March~25--30, 2012). IEEE, 2012. 5173--5176. DOI:~\href{https://doi.org/10.1109/ICASSP.2012.6289085}{10.1109/ICASSP.2012.6289085}.
	
	\bibitem{zymbler}
	\emph{Zymbler~M.} Best-match time series subsequence search on the Intel Many Integrated Core architecture // Proc. of the 19th East European Conf. on Advances in Databases and Information Systems (Poitiers, France, September~8--11, 2015). Lecture Notes in Computer Science. \textbf{9282}. Heidelberg: Springer, 2015. 275--286. DOI:~\href{https://doi.org/10.1007/978-3-319-23135-8\_19}{10.1007/978-3-319-23135-8\_19}.

\end{biblio}

\hfill
{\parbox[t]{3.7cm}{Поступила в редакцию\\04.12.2018}}
\vspace{1em}

\classify{} 

\title{The Use of MPI and OpenMP Technologies for Subsequence Similarity Search \\ in Very Large Time Series on Computer Cluster System \\with Nodes Based on the Intel Xeon Phi Knights Landing Many-core Processor}

\author{Ya.A.~Kraeva$^1$ and M.L.~Zymbler$^2$}

\addressen{%
	$^{1}$ South Ural State University, Department of System Programming; prospekt Lenina 76, Chelyabinsk, 454080, Russia; Master student, e-mail:  \href{mailto:kraevaya@susu.ru}{kraevaya@susu.ru}\\
	$^{2}$ South Ural State University, Department of System Programming; prospekt Lenina 76, Chelyabinsk, 454080, Russia; Cand. Sci., Associate Professor, e-mail: \href{mailto:mzym@susu.ru}{mzym@susu.ru}}


\received{December~4, 2018}

\maketitle{}

\begin{abstracten}
	Nowadays, subsequence similarity search is required in a wide range of time series mining applications: climate modeling, financial forecasts, medical research, etc. In most of these applications, the Dynamic Time Warping (DTW) similarity measure is used since DTW is empirically confirmed as one of the best similarity measure for most subject domains. Since the DTW measure has a quadratic computational complexity w.r.t. the length of query subsequence, a number of parallel algorithms for various many-core architectures have been developed, namely FPGA, GPU, and Intel MIC. In this article, we propose a new parallel algorithm for subsequence similarity search in very large time series on computer cluster systems with nodes based on Intel Xeon Phi Knights Landing (KNL) many-core processors. Computations are parallelized on two levels as follows: through MPI at the level of all cluster nodes, and through OpenMP within one cluster node. The algorithm involves additional data structures and redundant computations, which make it possible to effectively use the capabilities of vector computations on Phi KNL. Experimental evaluation of the algorithm on real-world and synthetic datasets shows that the proposed algorithm is highly scalable.
	
	\keywordsen{time series, similarity search, Dynamic Time Warping, parallel algorithm, OpenMP, Intel Xeon Phi, Knights Landing, data layout, vectorization.}
\end{abstracten}

\justifying

\begin{biblio_lat}
	
	
	\bibitem{VoevodinV02en}
	V.~V.~Voevodin, and Vl.~V.~Voevodin, \emph{The Parallel Computing}. (BHV-Petersburg, St.~Petersburg, 2002).
	
	
	
	\bibitem{DBLP:conf/sigmod/AthitsosPPKG08en}
	V.~Athitsos, P.~Papapetrou, M.~Potamias, et al., ``Approximate Embedding-Based Subsequence Matching of Time Series,'' in \textit{Proc. of the ACM SIGMOD Int. Conf. on Management of Data. Vancouver, Canada, June 10--12, 2008} (ACM, New York, 2008), pp.~365--378.  doi:~\href{https://dx.doi.org/10.1145/1376616.1376656}{10.1145/1376616.1376656}.
	
	\bibitem{baconen}
	D.~F.~Bacon, S.~L.~Graham, and O.~J.~Sharp, ``Compiler transformations for high-performance computing,'' ACM Comput. Surv. \textbf{26}~(4), 345--420 (1994).  doi:~\href{https://dx.doi.org/10.1145/197405.197406}{10.1145/197405.197406}.
	
	\bibitem{DBLP:conf/kdd/BerndtC94en}
	D.~J.~Berndt, and J.~Clifford, ``Using Dynamic Time Warping to Find Patterns in Time Series,'' in \textit{Proc. of the 1994 AAAI Workshop on Knowledge Discovery in Databases, Seattle, Washington, July, 1994} (AAAI Press, 1994), pp.~359--370.
	
	\bibitem{chrysosen}
	G.~Chrysos, ``Intel Xeon Phi Coprocessor (Codename Knights Corner),'' in \textit{2012 IEEE Hot Chips 24th Symposium (HCS), Cupertino, CA, USA, August~27--29, 2012} (IEEE, 2012), pp.~1--31. doi:~\href{https://dx.doi.org/10.1109/HOTCHIPS.2012.7476487}{10.1109/HOTCHIPS.2012.7476487}.
	
	\bibitem{DBLP:journals/pvldb/DingTSWK08en}
	H.~Ding, G.~Trajcevski, P.~Scheuermann, X.~Wang, and E.~Keogh, ``Querying and Mining of Time Series Data: Experimental Comparison of Representations and Distance Measures,'' Proc. of the VLDB Endowment. \textbf{1}~(2), 1542--1552 (2008).
	
	\bibitem{fuen}
	A.~W.~Fu, E.~J.~Keogh, L.~Y.~H.~Lau, C.~A.~Ratanamahatana, and R.~C.~Wong, ``Scaling and Time Warping in Time Series Querying,'' VLDB. \textbf{17}~(4), 899--921 (2008).  doi:~\href{https://doi.org/10.1007/s00778-006-0040-z}{10.1007/s00778-006-0040-z}.
	
	\bibitem{DBLP:conf/pvm/Gropp12en}
	W.~Gropp, ``MPI 3 and Beyond: Why MPI is Successful and What Challenges It Faces,'' in \textit{Proc. of the 19th European {MPI} Users' Group Meeting. Lecture Notes in Computer Science} (Springer, Berlin, 2012), Vol.~7490, pp.~1--9. doi:~\href{https://dx.doi.org/10.1007/978-3-642-33518-1\_1}{10.1007/978-3-642-33518-1\_1}.
	
	\bibitem{keoghen}
	E.~J.~Keogh, and C.~A.~Ratanamahatana, ``Exact Indexing of Dynamic Time Warping,'' Knowl. Inf. Syst. \textbf{7}~(3). pp.~358--386. (2005). doi:~\href{https://doi.org/10.1007/s10115-004-0154-9}{10.1007/s10115-004-0154-9}.
	
	
	\bibitem{KostenetskyS16en}
	P.~S.~Kostenetskiy, and A.~Y.~Safonov, ``SUSU Supercomputer Resources,'' in \textit{Proc. of the 10th Int. Conf. on Parallel Computing Technologies (PCT 2016), Arkhangelsk, Russia, March~29--31, 2016. CEUR Workshop Proc.} Vol.~1576, pp.~561--573.
	
	\bibitem{KraevaZ18en}
	Ya.~Kraeva, and M.~Zymbler ``An Efficient Subsequence Similarity Search on Modern Intel Many-core Processors for Data Intensive Applications,'' in \textit{Proc. of the 20th Int. Conf. on Data Analytics and Management in Data Intensive Domains, DAMDID/RCDL 2018, Moscow, Russia, October~9--12, 2018}, pp.~116--124.
	
	\bibitem{DBLP:books/bc/KumarGGK94en}
	V.~Kumar, A.~Grama, A.~Gupta, and G.~Karypis. \textit{Introduction to Parallel Computing} (Benjamin/Cummings, 1994).
	
	\bibitem{limen}
	S.~Lim, H.~Park, and S.~Kim, ``Using Multiple Indexes for Efficient Subsequence Matching in Time-Series Databases,'' in \textit{Proc. of the 11th Int. Conf. on Database Systems for Advanced Applications. Lecture Notes in Computer Science} (Springer, Berlin, 2006), Vol.~3882, pp.~65--79. doi:~\href{https://doi.org/10.1007/11733836\_7}{10.1007/11733836\_7}.
	
	\bibitem{DBLP:conf/sc/Mattson06en}
	T.~Mattson, ``Introduction to OpenMP,'' in \textit{Proc. of the 2006 ACM/IEEE Conf. on Supercomputing, Tampa, FL, USA, November~11--17, 2006} (ACM Press, New York, 2006), pp.~209.
	doi:~\href{https://dx.doi.org/10.1145/1188455.1188673}{10.1145/1188455.1188673}.
	
	\bibitem{DBLP:journals/cmpb/MazandaraniM18en}
	F.~N.~Mazandarani, and M.~Mohebbi, ``Wide Complex Tachycardia Discrimination Using Dynamic Time Warping of ECG Beats,'' Computer Methods and Programs in Biomedicine. \textbf{164}, 238--249 (2018). doi:~\href{https://dx.doi.org/10.1016/j.cmpb.2018.04.009}{10.1016/j.cmpb.2018.04.009}.
	
	\bibitem{MovchanZ16en}
	A.~V.~Movchan, and M.~L.~Zymbler, ``Parallel Implementation of Searching the Most Similar Subsequence in Time Series for Computer Systems with Distributed Memory,'' in \textit{Proc. of the 10th Annual International Scientific Conference on Parallel Computing Technologies (PCT 2016), Arkhangelsk, Russia, March 29--31, 2016. CEUR Workshop Proc.} Vol.~1576, pp.~615--628. 
	
	\bibitem{Pearson1905en}
	K.~Pearson, ``The Problem of the Random Walk,'' Nature. \textbf{72}~(1865). 294. (1905). doi:~\href{https://dx.doi.org/10.1038/072342a0}{10.1038/072342a0}.
	
	\bibitem{rakthanmanonen}
	T.~Rakthanmanon, B.~J.~L.~Campana, A.~Mueen, G.~E.~A.~P.~A.~Batista, M.~B.~Westover, et al., ``Searching and Mining Trillions of Time Series subsequences under Dynamic Time Warping,'' in \textit{Proc. of the 18th ACM SIGKDD Int. Conf. on Knowledge Discovery and Data Mining, Beijing, China, August~12--16, 2012} (ACM, New York, 2012), pp.~262--270. doi:~\href{http://doi.acm.org/10.1145/2339530.2339576}{10.1145/2339530.2339576}.
	
	\bibitem{Rebbapragada2009en}	
	U.~Rebbapragada, P.~Protopapas, C.~E.~Brodley, and C.~Alcock, ``Finding Anomalous Periodic Time Series,'' Machine Learning \textbf{74}~(3), 281--313 (2009). doi:~\href{https://dx.doi.org/10.1007/s10994-008-5093-3}{10.1007/s10994-008-5093-3}.	
	
	\bibitem{sakoeen}
	H.~Sakoe, and S.~Chiba, \textit{Readings in Speech Recognition.} (San Francisco, CA, USA, Morgan Kaufmann Publishers Inc., 1990), pp.~159--165.
	
	\bibitem{sakuraien}
	Y.~Sakurai, C.~Faloutsos, and M.~Yamamuro, ``Stream Monitoring under the Time Warping Distance,'' in \textit{Proc. of the 23rd Int. Conf. on Data Engineering, Istanbul, Turkey, April~15--20, 2007} (IEEE Computer Society, Washington, DC, USA, 2007). pp.~1046--1055. doi:~\href{https://dx.doi.org/10.1109/ICDE.2007.368963}{10.1109/ICDE.2007.368963}.	
	
	\bibitem{sarten}
	D.~Sart, A.~Mueen, W.~A.~Najjar, E.~J.~Keogh, and V.~Niennattrakul, ``Accelerating Dynamic Time Warping Subsequence Search with GPUs and FPGAs,'' in \textit{Proc. of the 2010 IEEE Int. Conf. on Data Mining, Sydney, Australia, December~14--17, 2010} (IEEE Computer Society, Washington, DC, USA, 2010), pp.~1001--1006. doi:~\href{https://doi.org/10.1109/ICDM.2010.21}{10.1109/ICDM.2010.21}.
	
	\bibitem{sodani}
	A.~Sodani, ``Knights Landing (KNL): 2nd generation Intel Xeon Phi processor,'' in \textit{2015 IEEE Hot Chips 27th Symposium (HCS), Cupertino, CA, USA, August~22--25, 2015} (IEEE, 2015), pp.~1--24.	
	
	\bibitem{SokolinskayaS16en}
	I.~Sokolinskaya, and L.~Sokolinsky, ``Revised Pursuit Algorithm for Solving Non-Stationary Linear Programming Problems on Modern Computing Clusters with Manycore Accelerators,'' Communications in Computer and Information Science. \textbf{687}, 212--223 (2016). doi:~\href{https://dx.doi.org/10.1007/978-3-319-55669-7\_17}{10.1007/978-3-319-55669-7\_17}.
	
	\bibitem{DBLP:conf/icacci/ShabibNNDPSATS15en}
	A.~Shabib, A.~Narang, C.~P.~Niddodi, et~al., ``Parallelization of Searching and Mining Time Series Data using Dynamic Time Warping,'' in \textit{2015 Int. Conf. on Advances in Computing, Communications and Informatics, Kochi, India, August 10--13, 2015} (IEEE, 2015), pp.~343–348. doi:~\href{https://dx.doi.org/10.1109/ICACCI.2015.7275633}{10.1109/ICACCI.2015.7275633}.
	
	\bibitem{srikanthanen}
	S.~Srikanthan, A.~Kumar, and R.~Gupta, ``Implementing the Dynamic Time Warping Algorithm in Multithreaded Environments for Real Time and Unsupervised Pattern Discovery,'' in \textit{2011 2nd International Conference on Computer and Communication Technology, Allahabad, India, September~15--17, 2011} (IEEE, 2015), pp.~394--398. doi:~\href{https://doi.org/10.1109/ICCCT.2011.6075111}{10.1109/ICCCT.2011.6075111}.
	
	\bibitem{takahashien}
	N.~Takahashi, T.~Yoshihisa, Y.~Sakurai, and M.~Kanazawa, ``A Parallelized Data Stream Processing System using Dynamic Time Warping Distance,'' in \textit{2009 Int. Conf. on Complex, Intelligent and Software Intensive Systems, Fukuoka, Japan, March~16--19, 2009} (IEEE, 2009), pp.~1100--1105. doi:~\href{https://doi.org/10.1109/CISIS.2009.77}{10.1109/CISIS.2009.77}.
	
	\bibitem{tarango}
	J.~Tarango, E.~J.~Keogh, and P.~Brisk, ``Instruction Set Extensions for Dynamic Time Warping'' in \textit{Proc. of the Int. Conf. on Hardware/Software Codesign and System Synthesis, Montreal, QC, Canada, September~29--October~4, 2013} (IEEE, 2013), pp.~18:1--18:10. doi:~\href{https://doi.org/10.1109/CODES-ISSS.2013.6659005}{10.1109/CODES-ISSS.2013.6659005}.
	
	\bibitem{wang}
	Z.~Wang, S.~Huang, L.~Wang, H.~Li, Y.~Wang, et~al., ``Accelerating Subsequence Similarity Search Based on Dynamic Time Warping Distance with FPGA'' in \textit{Proc. of the ACM/SIGDA International Symposium on Field Programmable Gate Arrays, Monterey, CA, USA, February~11--13, 2013} (ACM, New York, 2013), pp.~53--62. doi:~\href{http://doi.acm.org/10.1145/2435264.2435277}{10.1145/2435264.2435277}.
	
	\bibitem{zhang}
	Y.~Zhang, K.~Adl, and J.~R.~Glass, ``Fast Spoken Query Detection using
	Lower-Bound Dynamic Time Warping on Graphical Processing Units,'' in \textit{2012 IEEE Int. Conf. on Acoustics, Speech and Signal Proc., Kyoto, Japan, March~25--30, 2012} (IEEE, 2012), pp.~5173--5176. doi:~\href{https://doi.org/10.1109/ICASSP.2012.6289085}{10.1109/ICASSP.2012.6289085}.
	
	\bibitem{zymbler}
	M.~Zymbler, ``Best-Match Time Series Subsequence Search on the Intel Many Integrated Core Architecture'' in \textit{Proc. of the 19th East European Conf. on Advances in Databases and Information Systems, Poitiers, France, September~8--11, 2015. Lecture Notes in Computer Science} (Springer, Heidelberg, 2015), Vol.~9282, pp.~275--286. doi:~\href{https://doi.org/10.1007/978-3-319-23135-8\_19}{10.1007/978-3-319-23135-8\_19}.
	
\end{biblio_lat}

\end{document}